\documentclass{aa}

\usepackage{txfonts}

\begin{document} 

   \title{Planet and star synergy at high spectral resolution. \\A rationale for the characterisation of exoplanet atmospheres. }

   \subtitle{I. The Infrared}
\titlerunning{Planet and star synergy at high spectral resolution}

  \author{A. Chiavassa
          \inst{1}
          \and
          M. Brogi
          \inst{2,3,4}
          }

    \institute{Universit\'e C\^ote d'Azur, Observatoire de la C\^ote d'Azur, CNRS, Lagrange, CS 34229, Nice,  France \\
		\email{andrea.chiavassa@oca.eu}
         \and
 	 Department of Physics, University of Warwick, Coventry CV4 7AL, UK
         \and
         INAF - Osservatorio Astrofisico di Torino, Via Osservatorio 20, 10025, Pino Torinese, Italy
         \and
         Centre for Exoplanets and Habitability, University of Warwick, Gibbet Hill Road, Coventry CV4 7AL, UK
             }

   \date{...}

 
  \abstract
   {Spectroscopy of exoplanet atmospheres at high resolving powers is rapidly gaining popularity to measure the presence of atomic and molecular species. While this technique is particularly robust against contaminant absorption in the Earth's atmosphere, the non stationary stellar spectrum, in the form of either Doppler shift or distortion of the line profile during planetary transits, creates a non-negligible source of noise that can alter or even prevent detection.}
   {We aim at using state-of-the art three-dimensional stellar simulations to directly remove the signature of the star from observations, and prior to cross correlation with templates for the planet's atmosphere, commonly used to extract the faint exoplanet signal from noisy data.}
   {We compute synthetic spectra from 3D simulations of stellar convection resolved both spatially and temporally, and we couple them with an analytical model reproducing the correct geometry of a transiting exoplanet. We apply the method to the early K-dwarf, HD~189733, and re-analyse transmission and emission spectroscopy of its hosted exoplanet. In addition, we also analyse emission spectroscopy of the non transiting exoplanet 51~Pegasi~b, orbiting a solar-type star.}
   {We find a significant improvement in the planet detectability when removing the stellar spectrum with our method. In all cases, we show that the method is superior to a simple parametrisation of the stellar line profile or to the use of one-dimensional stellar models. We show that this is due to the intrinsic treatment of convection in 3D simulations, which allows us to correctly reproduce asymmetric and/or blue-shifted spectral lines, and intrinsically model center-to-limb variation and Rossiter-McLaughlin effect potentially altering the interpretation of exoplanet transmission spectra. In the case of 51~Pegasi~b, we succeed at confirming a previous tentative detection of the planet's $K$-band spectrum due to the improved suppression of stellar residuals.}
   {Future high-resolution observations will benefit from the synergy with stellar spectroscopy, and can be used to test the correct modelling of physical processes in stellar atmospheres. We highlight key improvements in modelling techniques and knowledge of opacity sources to extend this work to shorter wavelengths and later-type stars.}

   \maketitle

%

\section{Introduction}

The remote atmospheric characterisation of planets outside our solar system (exoplanets) is considered a key milestone to unravel their physical and chemical composition \citep{2009ApJ...690.1056M}, their formation scenarios \citep{2012ApJ...758...36M, 2015ApJ...815..109P, 2018A&A...613A..14E}, and ultimately the presence of conditions amenable to life \citep{2018AsBio..18..663S}. Albeit challenging, these observations have reached a level of maturity where multiple observing techniques can be applied to both space- and ground-based telescopes. Among these, high-resolution spectroscopy (HRS) at resolving powers $R>25,000$ led to the detection of molecular (CO, H$_2$O, CH$_4$, HCN, TiO) and atomic (H, He, K, Na, Mg, Fe, Ti) species in a dozen exoplanets \citep[see e.g.][and references therein, for a recent review]{birkby_review}. There are two unique aspects of HRS that make it particularly suitable for exoplanet characterisation. Firstly, at high resolving powers molecular species are partially resolved into a dense forest of lines that can be robustly identified by line-matching techniques such as cross-correlation. Secondly, the planet is subject to a detectable Doppler shift as it moves along the orbit, which can be used to solve for the orbital inclination of non-transiting systems similarly to spectroscopic binary stars \citep[e.g.][]{2012Natur.486..502B}. Doppler shift and broadening also constrain atmospheric winds \citep{Louden2015, 2016ApJ...817..106B, 2019AJ....157..209F} and overall bulk planet rotation \citep{2014Natur.509...63S, Schwarz2016}. 

Initially limited to the brightest stars, HRS has recently achieved a substantial increase in sensitivity due to the advent of new, performing infrared spectrographs mounted at large and medium-size telescope facilities, among which GIANO \citep{giano2014}, CARMENES \citep{carmenes2014}, SPIRou \citep{spirou2014}, and soon CRIRES+ \citep{criresplus2014}. Coupled with the progressive discovery of more planets orbiting bright stars, either from radial-velocity or transit surveys \citep[e.g.][]{hd219134,pimensae}, this means that the sample of exoplanets potentially detectable with HRS is now on the order of dozens, and will likely be of the order of hundreds after the end of the TESS mission \citep{2018ApJS..239....2B}. Importantly for future studies, it has been shown that HRS with ground-based facilities and space-borne, low-resolution spectroscopy are highly complementary and can be combined \citep{2017ApJ...839L...2B, 2019AJ....157..114B} to improve contraints on the chemical make up and thermal structure of atmospheres. 

Owing to the ability to disentangle sources with different Doppler signature, HRS is particularly suitable to diagnose potential sources of spurious signals, which can severely complicate the interpretation of exoplanet spectra. One of these is the non-uniformity of the planet-hosting stars. 
Their photosphere is covered with a complex and stochastic pattern associated with convective heat transport (i.e., granulation). Convection manifests in the surface layers as a particular pattern of downflowing cooler plasma and bright areas where hot plasma rises \citep{2009LRSP....6....2N}. Convection is a difficult process to understand because it is non-local, and three-dimensional, and it involves nonlinear interactions over many disparate length scales. In this context, the use of numerical three-dimensional (3D) radiative hydrodynamical (RHD) simulations of stellar convection is crucial, but has only become possible in recent years with the increase of computational power, resulting in large grids of simulations covering a substantial portion of the Hertzsprung-Russell diagram \citep{2013A&A...557A..26M,2013ApJ...769...18T,2013A&A...558A..48B,2009MmSAI..80..711L}. The use of those simulations has proven that the convection-related surface structures have different size, depth, and temporal variations, depending on the stellar type \citep{2013A&A...557A...7T,2013A&A...558A..49B,2014arXiv1405.7628M}.  More importantly, the related activity (in addition to other phenomena such as magnetic spots, rotation, dust, etc.) have an impact in stellar parameters determination \citep{2011A&A...534L...3B,2012A&A...545A..17C, 2012A&A...540A...5C}, radial velocity  \citep{2008sf2a.conf....3B,2011JPhCS.328a2012C,2013A&A...550A.103A},  chemical abundances determinations \citep{2005ASPC..336...25A,2009ARA&A..47..481A,2011SoPh..268..255C}, photometric colours \citep{2018A&A...611A..11C,2017MmSAI..88...90B},  and on planet detection \citep{2015A&A...573A..90M,2017A&A...597A..94C}. 

In the context of exoplanet studies, the use of 3D RHD simulations has already been endavoured in several works. \cite{2015A&A...576A..13C} showed that 3D simulations are better suited than ad-hoc limb-darkened laws for the interpretation of transit light curves in terms of ingress/egress slopes as well as the emerging flux. \cite{2017A&A...597A..94C} continued this work evaluating the impact of granulation from optical to IR wavelengths for several planet/star systems and determined the granulation error budget on the determination of the planetary radius. \cite{2016A&A...588A.127C} studied the impact of the stellar surface on the Rossiter-Mclaughlin effect, differential rotation, and the resulting star-planet alignment. \cite{2018ApJ...866...55C} inspected the impact of the granulation on the resultant line profiles and found induced center-to-limb variations in shape and net position. \cite{2017A&A...605A..90D,2018A&A...616A.144D} scrutinised the possibility to study stellar microvariability highlighting small surface segments, hidden behind the planet transits, with high spectral resolution observations.

In this work we demonstrate that 3D RHD simulations can be already applied to correct existing HRS observations of exoplanets and lead to a noticeable improvement in the detectability of their atmospheres compared to uncorrected spectra, or spectra corrected with 1D stellar models. Section~\ref{sec:3dmodels} introduces the numerical methods used, Section~\ref{sec:stellarvar} evidence the stellar intrinsic spatial and temporal variability, Section~\ref{sec:obs} explains the observational tools developed, Sections~\ref{hd189} and~\ref{51peg} report two direct applications of our approach.

\section{Three-dimensional radiative-hydrodynamical approach}\label{sec:3dmodels}

\subsection{Simulations}

The stellar surface convection simulations used in this work are calculated using the {\sc Stagger Code} \citep{2011A&A...528A..32C,2009LRSP....6....2N} that is a state-of-the-art (magneto)hydrodynamic code that solves the time-dependent hydrodynamic equations for mass, momentum, and energy conservation, coupled with the 3D radiative transfer equation in order to account correctly for the interaction between the radiation field and the plasma. The equations are solved on a staggered mesh where the thermodynamical scalar variables (density, internal energy, and temperature) are cell centered, while the fluxes are defined on the cell faces. This scheme has several numerical advantages when simulating surface
convection. It is robust against shocks and ensures conservation of the thermodynamic variables.  The domain of the simulation contains an entropy minimum at the surface that is sufficiently deep to ensure a flat entropy profile at the bottom. The code uses periodic boundary conditions horizontally and open boundaries
vertically. At the bottom of the simulation, the inflows have a constant entropy and
pressure. The outflows are not tightly constrained and are free to pass through the
boundary. The code is based on a sixth-order explicit finite-difference scheme and a, fifth-order interpolation. It employs realistic input physics: the equation of state is an updated version of the one described by \cite{1988ApJ...331..815M}, and the radiative transfer is calculated for a large number over wavelength points merged into 12 opacity bins \citep{1982A&A...107....1N,2000ApJ...536..465S,2013A&A...557A..26M}. They include continuous absorption opacities and scattering coefficients from \cite{2010A&A...517A..49H} as well as line opacities described in \cite{2008A&A...486..951G}, which in turn are based on the VALD-2 database \citep{2001ASPC..223..878S} of atomic lines. A solar chemical composition is assumed \citep{asplund09}.

\begin{table*}
\centering
\caption{3D RHD simulations from \textsc{Stagger}-grid \citep{2013A&A...557A..26M} used in this work for the observed systems reported in last column. $T_{\rm{eff}}$ are from \cite{2018A&A...611A..11C}.}             
\label{simus}      
\centering                          
\renewcommand{\footnoterule}{} 
\begin{tabular}{c c c c c c}        
\hline\hline                 
$T_{\rm{eff}}$ & [Fe/H]  & $\log g$ & $x,y,z$-dimensions & $x,y,z$-resolution   & Used for the\\
$[\rm{K}]$ & & [cgs]  & [Mm]  & [grid points] & observed system  \\
\hline
5788 (solar) & 0.0 & 4.44 &  8.00$\times$8.00$\times$4.97 & 240$\times$240$\times$230 & 51~Pegasi~b \\
4982 (K~dwarf) & 0.0 & 4.50 & 5.00$\times$5.00$\times$4.24 & 240$\times$240$\times$230 &  HD~189733~b \\
\hline\hline                          
\end{tabular}
\end{table*}

\subsection{Synthetic spectra in the infrared region}

We used the 3D pure-LTE radiative transfer code \textsc{Optim3D} \citep{2009A&A...506.1351C} to compute synthetic spectra from the snapshots of the RHD simulations reported in Table~\ref{simus} and taken from the \textsc{Stagger}-grid \citep{2013A&A...557A..26M}. The code takes into account the Doppler shifts due to convective motions. The radiative transfer equation is solved monochromatically using pre-tabulated extinction coefficients as a function of temperature, density, and wavelength. The lookup tables were computed for the same chemical compositions as the RHD simulations using the same extensive atomic and molecular continuum and line opacity data as the latest generation of MARCS models \citep{2008A&A...486..951G}. We assumed microturbulence equal to zero since the velocity fields inherent in RHD simulations are expected to self-consistently and adequately account for non-thermal Doppler broadening of spectral lines \citep{2000A&A...359..755A}. More details about \textsc{Optim3D} can be found in \citet{2009A&A...506.1351C,2010A&A...524A..93C}.

\begin{figure}
	\centering
		\includegraphics[width=1.\hsize]{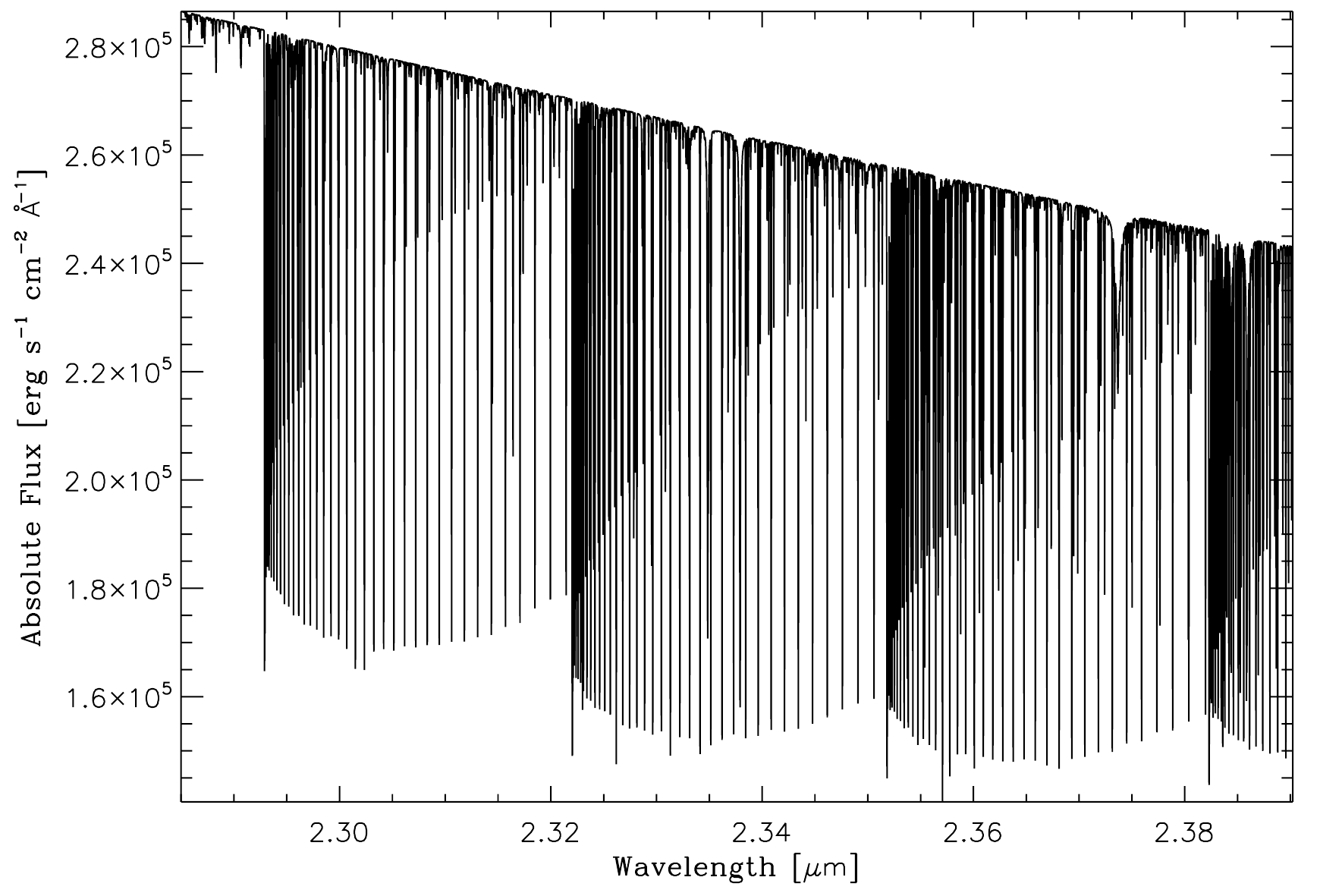}   
	\caption{Example of synthetic spectrum from the K-dwarf RHD simulation of Table~\ref{simus}. The spectral lines correspond largerly to 2-0 rovibrational band of carbon monoxide.} 
	\label{full_range_CO}
\end{figure}

As in \cite{2018A&A...611A..11C}, we computed spectra with a constant resolving power of $\lambda/\Delta\lambda$ = 300\,000 from 22\,850 to 23\,900 \AA, casting vertical rays through the computational box of the RHD simulations for all required wavelengths, for ten box tilting angles $\mu=\cos(\theta)$ = [1.00, 0.90, 0.80, 0.70, 0.50, 0.30, 0.20, 0.10, 0.05, 0.01], where $\theta$ is the angle between the normal to the surface and the line of sight, and four azimuths rotations $\phi$ = [0$^\circ$, 90$^\circ$, 180$^\circ$, 270$^\circ$]. In addition, a temporal average is also performed over ten snapshots adequately spaced  so as to capture several convective turnovers. The final resulting spectra are:

\begin{itemize}
\item a temporally averaged intensity spectrum at different $\mu$ and $\phi$-angles, which will be used to model the changes in the disk-averaged stellar spectrum during the transit of exoplanet HD~189\,733\,b (Sec.~\ref{hd189_tr}).
\item a temporally and spatially averaged flux (Fig.~\ref{full_range_CO}), which will be used to correct the stellar spectra in datasets targeting the thermal emission of exoplanets (Sections~\ref{hd189_day} and \ref{51peg}).
\end{itemize}

\begin{figure}
   \centering
    \includegraphics[width=1.\hsize]{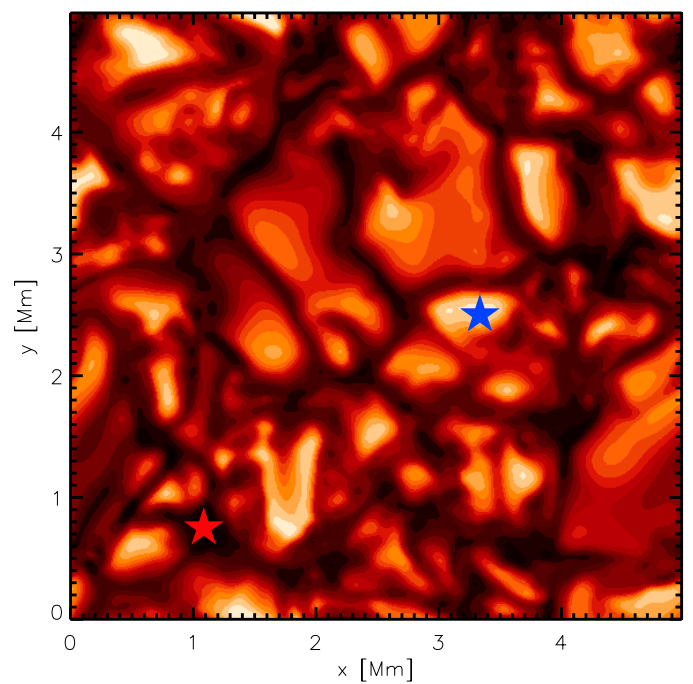}
      \caption{Spatially resolved profiles at disk centre ($\mu$=1.0, top panel) of one CO line ($\lambda$=23015.002 \AA, log $gf$=0.221, $\chi$=-5.474 eV) across the granulation pattern (bottom panel) of a K-dwarf simulation (Table~\ref{simus}). The solid red (intergranular lane) and blue (granule) lines displays two particular positions (colored star symbols) extracted from the intensity map.}
        \label{multiple_lines}
   \end{figure}

\section{Intrinsic stellar variability and its impact in the infrared}\label{sec:stellarvar}

The granulation pattern is associated with heat transport by convection. The bright areas on the stellar surfaces, the granules (Fig.~\ref{multiple_lines}, bottom panel), are the locations of upflowing hot plasma, while the dark intergranular lanes are the locations of downflowing cooler plasma. Additionally, the horizontal scale on which radiative cooling drives the convective motions is linked with the granulation diameter \citep{2009LRSP....6....2N}. One piece of evidence of the convective related surface structures comes from the unresolved spectral lines, in particular at high spectral resolution. In fact, they combine important properties such as velocity amplitudes and velocity-intensity correlations, which affect the line shape, shift, and asymmetries. These structures derive from the convective flows in the solar photosphere and solar oscillations \citep{2000A&A...359..669A, 2009LRSP....6....2N}.

\begin{figure}
	\centering
		\includegraphics[width=1.\hsize]{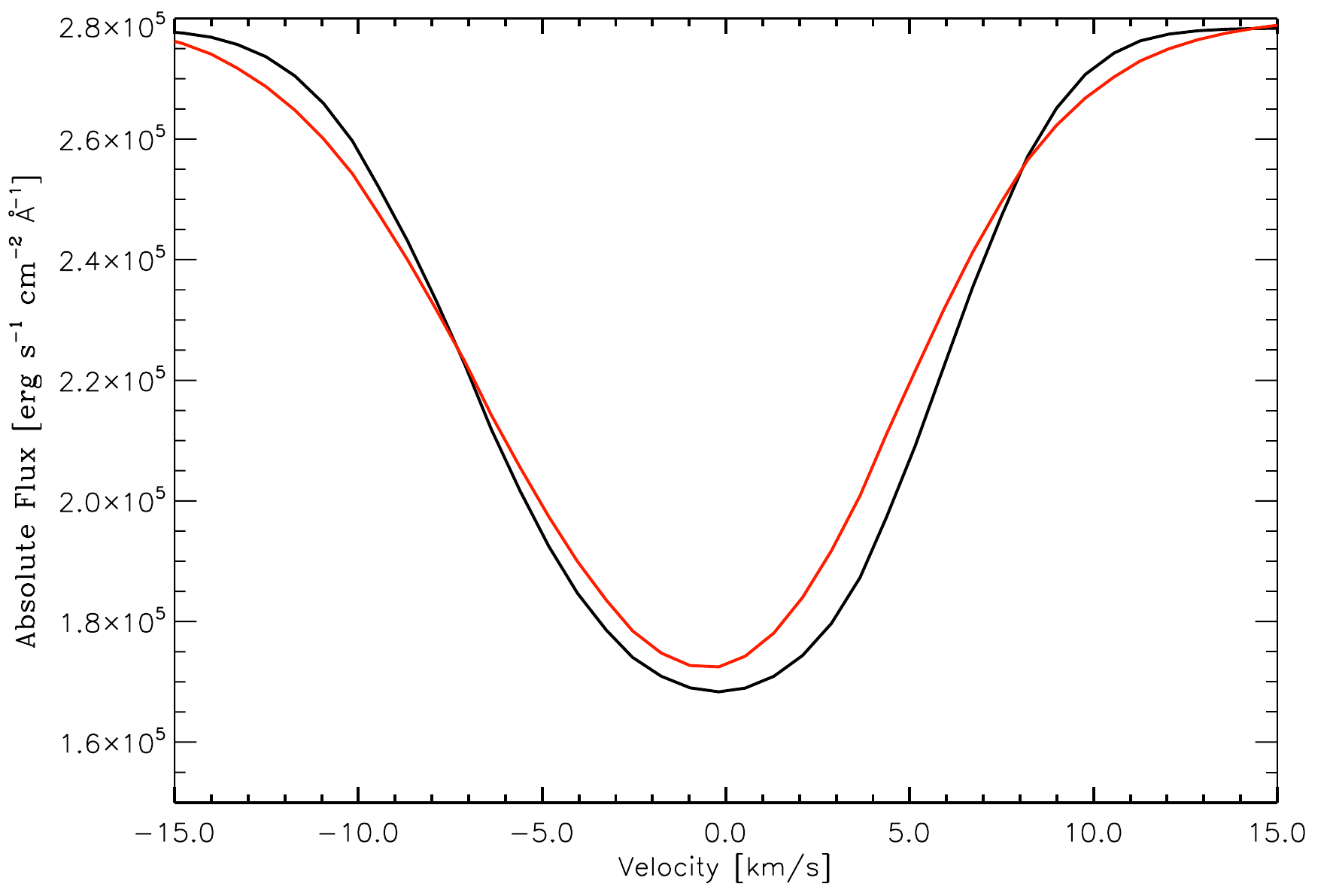}   
	\caption{Spatial and temporal average spectrum (red) of one CO line (same as in Fig.~\ref{multiple_lines}) for K-dwarf simulation  (Table~\ref{simus}) together with a 1D hydrostatic MARCS model \citep{2008A&A...486..951G} with the same stellar and spectral line parameters.} 
	\label{1D_vs_3D}
\end{figure}

Fig.~\ref{multiple_lines} displays an example of the spatially resolved intensities corresponding to different regions across the granulation pattern. Correlations of velocity and temperature cause characteristic asymmetries of spectral lines as well as net blue- or red-shifts depending on the area probed \citep{dravins87,2005oasp.book.....G}. Bright and rising convective elements (granule) being blue-shifted and contributing more photons than the cool dark shrinking gas (intergranular lanes) that are red-shifted \citep{1982ARA&A..20...61D}. As a consequence, the resulting spatially averaged absorption line appears blue-shifted as in Fig.~\ref{1D_vs_3D} (red line).

An evident and important difference when using RHD simulation with respect to traditional one-dimensional (1D), stationary, hydrostatic model stellar atmospheres such as MARCS \citep{2008A&A...486..951G}, ATLAS \citep{2005MSAIS...8...14K}, or PHOENIX \citep{2013A&A...553A...6H} is the treatment of convection. 1D models can treat convective energy transport in an approximate manners \citep[e.g., mixing-length theory, ][]{1958ZA.....46..108B} that are all dependent on a number of free parameters. However, since the convection zone reaches up to the optical surface in main sequence stars, convection directly influences the spectrum formation both by modifying the mean stratification and by introducing
inhomogeneities and velocity fields in the photosphere. It is therefore important to properly account for their effects in order to extract accurate and precise information from the analysis of stellar spectra. This is displayed in Fig.~\ref{1D_vs_3D} where the shape of CO line is entirely symmetric and centered to zero km/s in the case of 1D model (black line) and 
blue-shifted in the case of RHD simulation (red line). Each line has unique fingerprints in the spectrum that depends on line strength, depth, shift, width, and asymmetry across the granulation pattern depending on their height of formation and sensitivity to the atmospheric conditions. In this context, the line strength of lines  plays a major role \citep{2000A&A...359..743A}.

\begin{figure}
	\centering
		\includegraphics[width=1.\hsize]{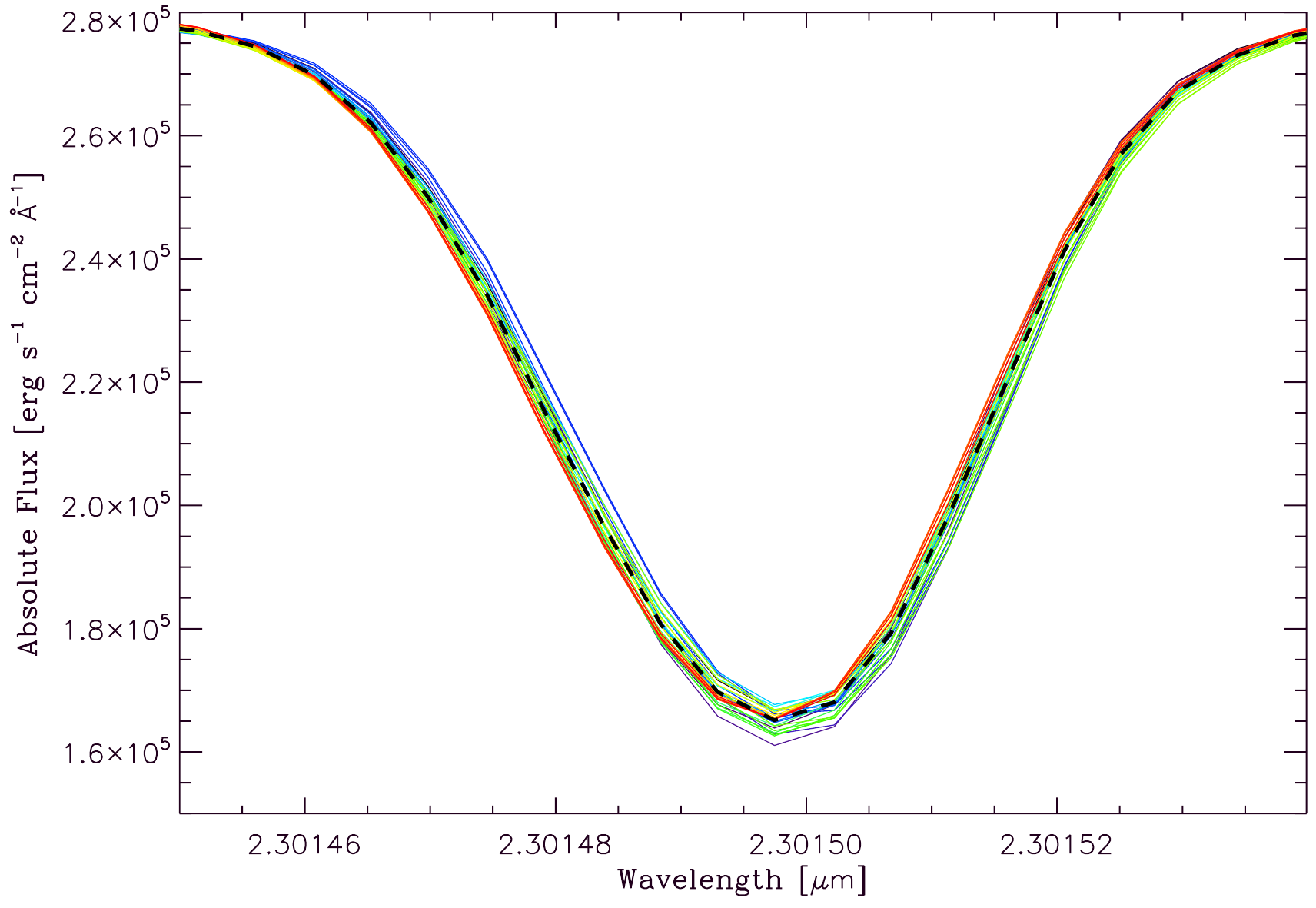}    
	\caption{Temporal variation of spatial averaged spectral CO line (same as in Fig.~\ref{stars_CO_lines}) for K-dwarf simulation (Table~\ref{simus}). The increasing time changes with colors from blue to red (ten snapshots and 70 minutes spanned in time)} 
	\label{CO_line_variation}
\end{figure}

Eventually, one last important point of concern is the temporal variation of spatial averaged spectral lines during planet transits. Fig.~\ref{CO_line_variation} displays the uncorrelated temporal variations over slightly more than one hour of simulation time. This variability has already been detected in several occasions. For example, it has been observed that the Sun’s total irradiance varies on all timescales relevant for transit surveys, from minutes to months \citep{2004A&A...414.1139A}. Moreover, granulation manifested on photometric variability of the SOHO quiet-Sun data \citep[ranging between 10 to 50 ppm][]{2002ApJ...575..493J,1997SoPh..170....1F}, which have been explained by \cite{2017A&A...597A..94C} with the same RHD simulations used in this work. 

In next Sections, we apply synthetic spectra computed for 3D RHD simulations to correct existing HRS observations of exoplanets and check the improvement in the detectability of their atmospheres.

\section{Observations and data analysis}\label{sec:obs}

\subsection{Observations}
\begin{figure*}[ht!]
	\centering
		\includegraphics[width=1.0\hsize]{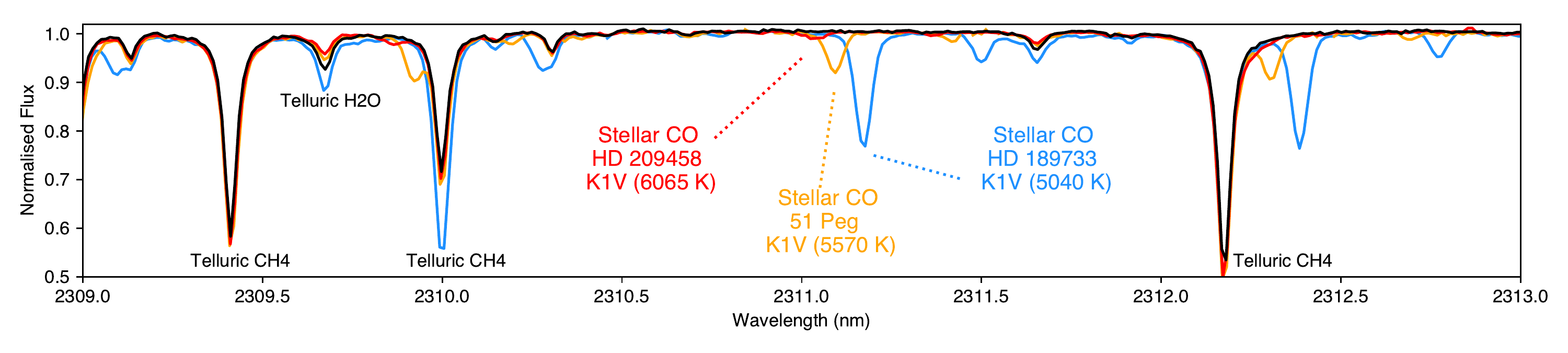}  
		\vspace{-0.5cm}
	\caption{Example CRIRES spectra in the wavelength range 2309-2313 nm, color coded according to the stellar host: $\tau$ Bo\"otis in black \citep{2012Natur.486..502B}, HD 209458 in red \citep{2017ApJ...839L...2B}, 51~Pegasi in gold \citep{brogi13}, and HD 189733 in blue \citep{2016ApJ...817..106B}. The amplitude of stellar CO lines increases with decreasing effective temperature as expected, while their position varies due to the different barycentric + systemic velocity of the targets for different targets and/or night of observations. In contrast, lines formed in the Earth's atmosphere (telluric lines) are stationary in wavelength and only slightly vary in depth due to airmass effects.}
	\label{stars_CO_lines}
\end{figure*}

We applied the models described in the previous section to correct the stellar spectrum imprinted on real spectroscopic data of bright exoplanet host stars. We devote this section to detail the instrumental setup, in common between data sets, and the basic data analysis applied to the spectra. The spectra were acquired with the Cryogenic Infrared Echelle Spectrograph \citep[CRIRES,][]{crires} mounted at the Nasmyth-A focus of the 8.2-m ESO Very Large Telescope UT1. The spectrograph was set to observe with a relatively narrow slit of $0.2^{\prime\prime}$ in order to achieve the maximum resolving power of $R \approx 100,000$. A standard observing sequence with ABBA nodding pattern was used for these observations, with the telescope slewing by 10$^{\prime\prime}$ along the slit between nodding positions A and B. Difference images (A-B or B-A) were therefore used to subtract out the contribution from sky emission lines and thermal background. One peculiar aspect of these observations is that the temporal information is preserved, i.e. each couple of A-B or B-A frames is combined and the corresponding one-dimensional spectrum extracted, which differs from the common practice of co-adding all the spectra obtained during an observing night. This strategy allows us to have a time resolution between 1 and 2 minutes per combined exposure, enough to prevent the change in orbital radial velocity of exoplanets from broadening the planet spectral lines.

Extraction of the one-dimensional spectra is obtained via the standard ESO pipeline. Each dataset described below has been published at different stages of the lifetime of CRIRES, and therefore different pipeline versions have been used. The data of 51~Pegasi \citep{brogi13} and the thermal emission data of HD 189733 \citep{dekok13} were processed with version 1.11.0 of the pipeline. The transit data of HD 189733 were instead processed with version 2.1.3 \citep{2016ApJ...817..106B} of the pipeline.

CRIRES images spectra on four detectors physically separated by a few mm. This means that, although our data covers the interval between 2.27 and 2.35 $\mu$m, there are corresponding small gaps in the spectral coverage. Our data target most of the 2-0 rovibrational band of carbon monoxide (CO). In high-temperature exoplanet spectra, water vapour is also expected to produce significant spectral lines and had indeed been reported multiple times. Methane could also contribute with a dense forest of weaker lines, but so far its presence has not been confirmed at these wavelengths. 

Interestingly, since the same CO lines are also present in the spectrum of the two stellar hosts examined in this paper, there is a chance of contaminating the planet signal with uncorrected stellar lines when the planet-star differential radial velocity is small. For planets on circular orbits, this happens during transit but also close to superior conjunction. A small fraction of a CRIRES spectrum is shown in Fig.~\ref{stars_CO_lines}, with telluric lines - mostly due to methane in this case - absorbing at a fixed wavelength and with approximately constant intensity, while the stellar CO lines vary in position and intensity according to the stellar systemic velocity and spectral type.

\subsection{Data calibration pipeline}
To enable accurate removal of the stellar spectrum, not only do we need to obtain a reliable wavelength solution, but also to measure the instrumental profile of CRIRES for each individual spectrum. In preparation for these steps, we use the ESO Sky Model calculator to precompute a coarse grid of telluric models in airmass (in the range 1.4 to 2.9, enough to cover the observations) and precipitable water vapour (PWV, in the range 0.5 to 3.5 mm). 

The pixel-wavelength solution of CRIRES is known to drift by up to a pixel (approximately 0.1 \AA\ at the wavelength of these observations) during a night. This means that arc frames taken at the end of the night are insufficient to provide a stable and accurate solution. The position and known wavelength of telluric lines imprinted on each spectrum is instead used as simultaneous reference to solve for the pixel-wavelength solution. If spectral lines from the parent star are also prominent, these can lower the quality of the solution due to blends and to the non-negligible change in the barycentric velocity of the observer during the observations. Although stellar spectral lines have been neglected in previous literature, in this work we revise the calibration pipeline in order to include the effects of a variable (in Doppler space) stellar spectrum. We proceed as follows:
\begin{enumerate}

\item We start from a guess wavelength solution which is the solution published in the literature for the three sets of data. As in past work, we assume a quadratic pixel-wavelength solution for each of the CRIRES detectors. This means that any wavelength solutions can be completely described by a triplet of points $(x_1,\lambda_1), (x_2,\lambda_2), (x_3,\lambda_3)$, where $x$ is the pixel position and $\lambda$ the corresponding wavelength. We adopt $(x_1, x_2, x_3) = (268, 512, 768)$ to evenly sample the 1024 pixels of a CRIRES detector. 

\item To assess the goodness of the wavelength solution, we compare the position of spectral lines in the observed spectra to those in model spectra via cross correlation. As results are almost exclusively dependent on line position, we do not need to exactly match the amplitude of spectral lines at this stage. Therefore, we choose a telluric model as close as possible to the average airmass of each night of observations, and with PWV=1.5 mm, the median value at Cerro Paranal.

\item We launch a Markov-Chain Monte Carlo (MCMC) using the Python package \texttt{emcee} \citep{foreman13} where we vary the $\lambda_i$ (3 parameters) at each step of the MonteCarlo. We interpolate the telluric and stellar models (the latter Doppler-shifted accounting for the systemic, barycentric, and orbital radial velocity) to the tested solution,  cross correlate the combined model with the data, and translate the cross correlation value into log-likelihood as in \citet{zucker03} to drive the MCMC. 

\item We re-grid the data to a common wavelength solution with constant steps in $\Delta\lambda/\lambda$. This is necessary to correctly estimate the instrumental profile, as explained below.

\end{enumerate}

\begin{figure}
	\centering
	\includegraphics[width=1.\hsize]{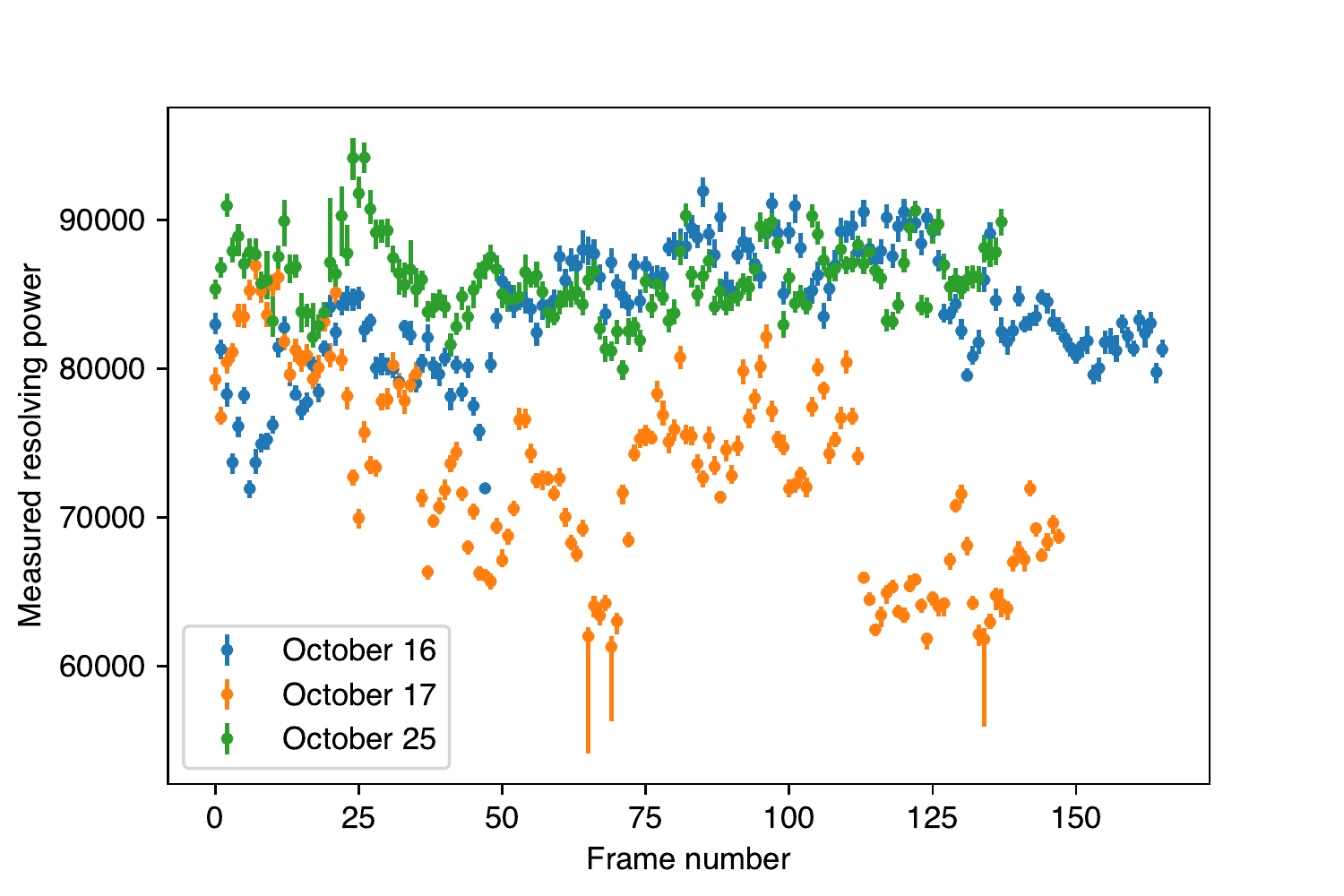}\\
	\vspace{-0.45cm} 
	\includegraphics[width=1.\hsize]{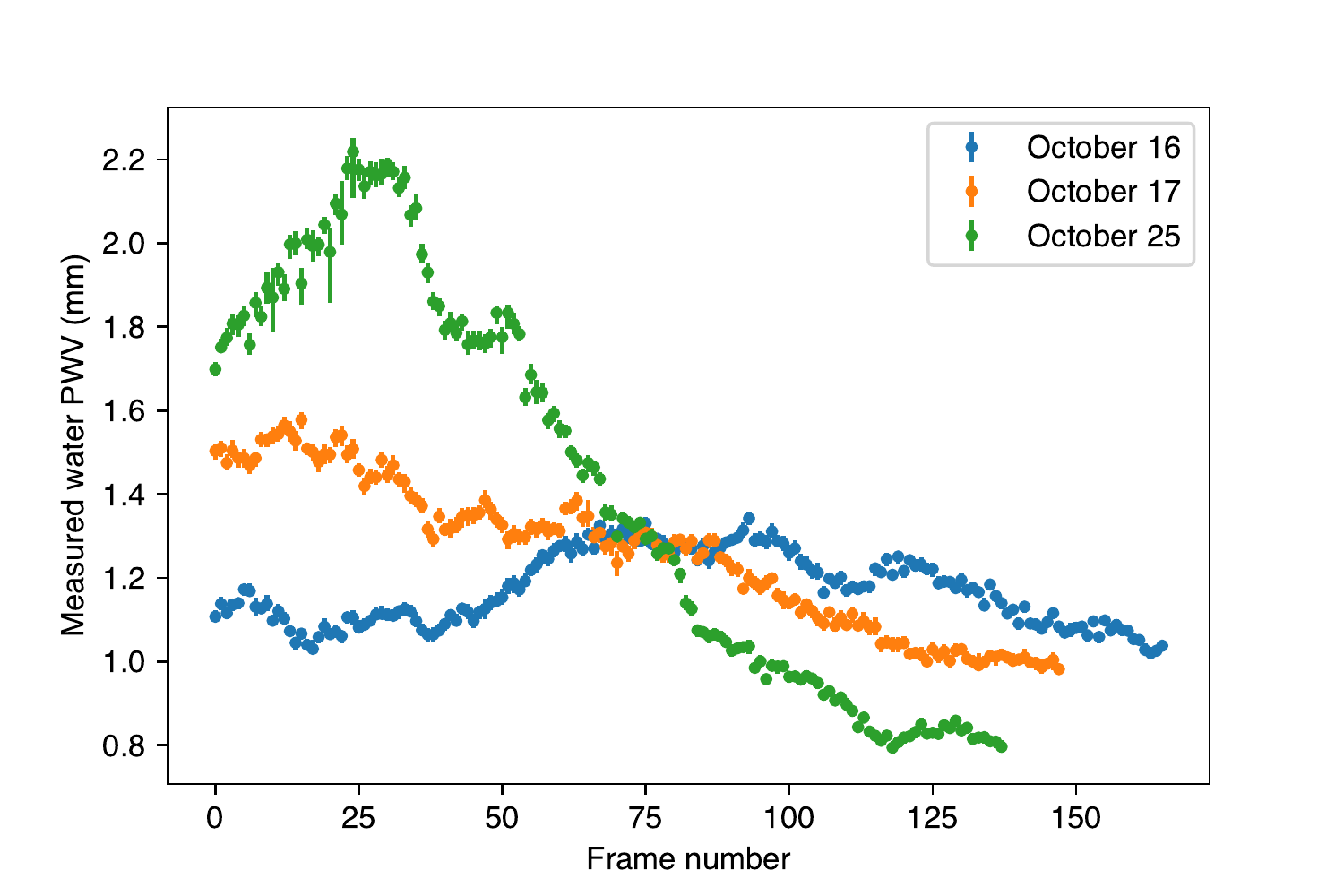}
	\caption{Fitted spectral resolving power (top panel) and precipitable water vapour (bottom panel) in the three nights of observations of 51 Pegasi b (see Sect. \ref{51peg}). It demonstrates the diagnostic capabilities of our automated data analysis pipeline to extract key information about instrumental and weather conditions during the observations.}
	\label{fig:51peg_parameters}
\end{figure}

The use of a MonteCarlo technique allows us to robustly assess the precision of the wavelength calibration. Each of the $\lambda_i$ can be determined with a typical error between $5\times10^{-4}$ and $10^{-3}$ nm at these wavelengths, which is 4-8\% of the size of a CRIRES pixel.

We approximate the Instrumental Profile (IP) with a Gaussian profile with variable FWHM. The width of the profile depends on the changing resolving power of the instrument due to e.g. varying slit illumination, pointing jitters, pressure and temperature changes, and any inaccuracies in combining A and B frames. Accurate IP estimation requires an unbroadened model as close as possible to the actual data. This is why at this stage we also fit for airmass and PWV to adjust the depth of the telluric spectrum to the observations. The stellar spectrum is further broadened with a rotational kernel corresponding to the literature values of the stellar projected rotational velocity, or $v\sin(i)$. Attempts to fit for $v\sin(i)$ as a free parameter resulted in no meaningful constraints within $\pm1$ km s$^{-1}$. 

We launch an MCMC on each observed spectrum where airmass, resolving power, and PWV are fitted at once (3 parameters). As for the wavelength calibration, the procedure uses the cross-correlation-to-likelihood mapping of \citet{zucker03}. Each telluric model spectrum is now obtained by bi-linearly interpolating across the pre-computed grid of models in log-space, and both telluric and stellar spectrum are broadened by the variable IP prior to cross correlation. Fig.~\ref{fig:51peg_parameters} shows the typical behaviour of resolving power and PWV over three different night of observations, with error bars reporting the 1-$\sigma$ confidence intervals from the MCMC.


\begin{figure*}
	\centering
		\includegraphics[width=1.\hsize]{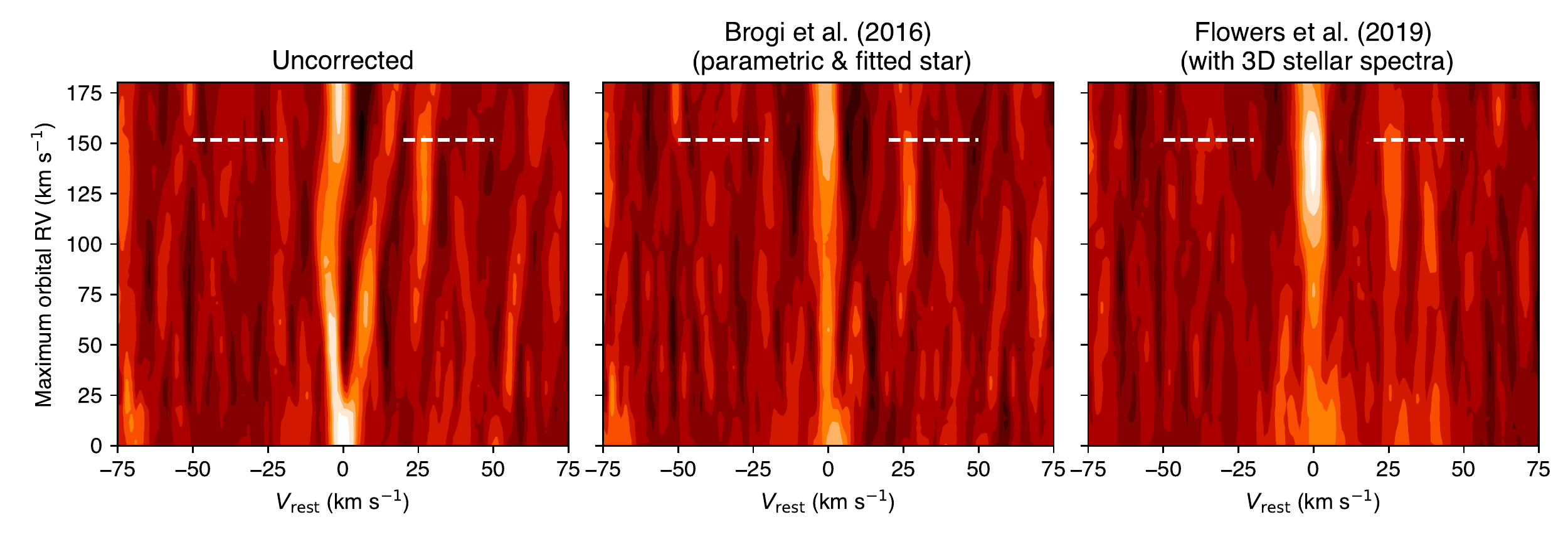}    
		\vspace{-0.6cm}
	\caption{Total cross-correlation signal from carbon monoxide in the transmission spectrum of HD 189733 b, as function of stellar removal applied to the data. Left panel: uncorrected spectra show a candidate signal at the expected planet maximum orbital velocity ($K_\mathrm{P} \sim$ 151 km s$^{-1}$, white dashed lines), but also significant stellar contamination at all values of $K_\mathrm{P}$ from uncorrected CO stellar lines. Middle panel: Application of a parametric 1D stellar model (middle panel) partially mitigate the problem and allows recovery of the CO signal at 5$\sigma$ \citep{2016ApJ...817..106B}. There is however still residual stellar signal that hinders the detection. Right panel: application of the phase-resolved 3D stellar models accounting for the geometry of transit \citep[this work and][]{2019AJ....157..209F} allows us to unambiguosly detect the planet CO signal at $>7\sigma$ of confidence level.}
	\label{fig:hd189_transmission}
\end{figure*}

At the end of the calibration process, the final wavelength solution and measured IP are used to remove the stellar spectrum from the data. This is done by shifting the appropriate stellar model (see details in each individual Section below) based on the three radial-velocity components listed above (systemic, barycentric, gravitational). The shifted model is then broadened by convolution with the IP and divided out from the spectra. Since the stellar model had its continuum previously normalised, no re-scaling is necessary before dividing it out.

\begin{figure}
	\centering
	\includegraphics[width=1.0\hsize]{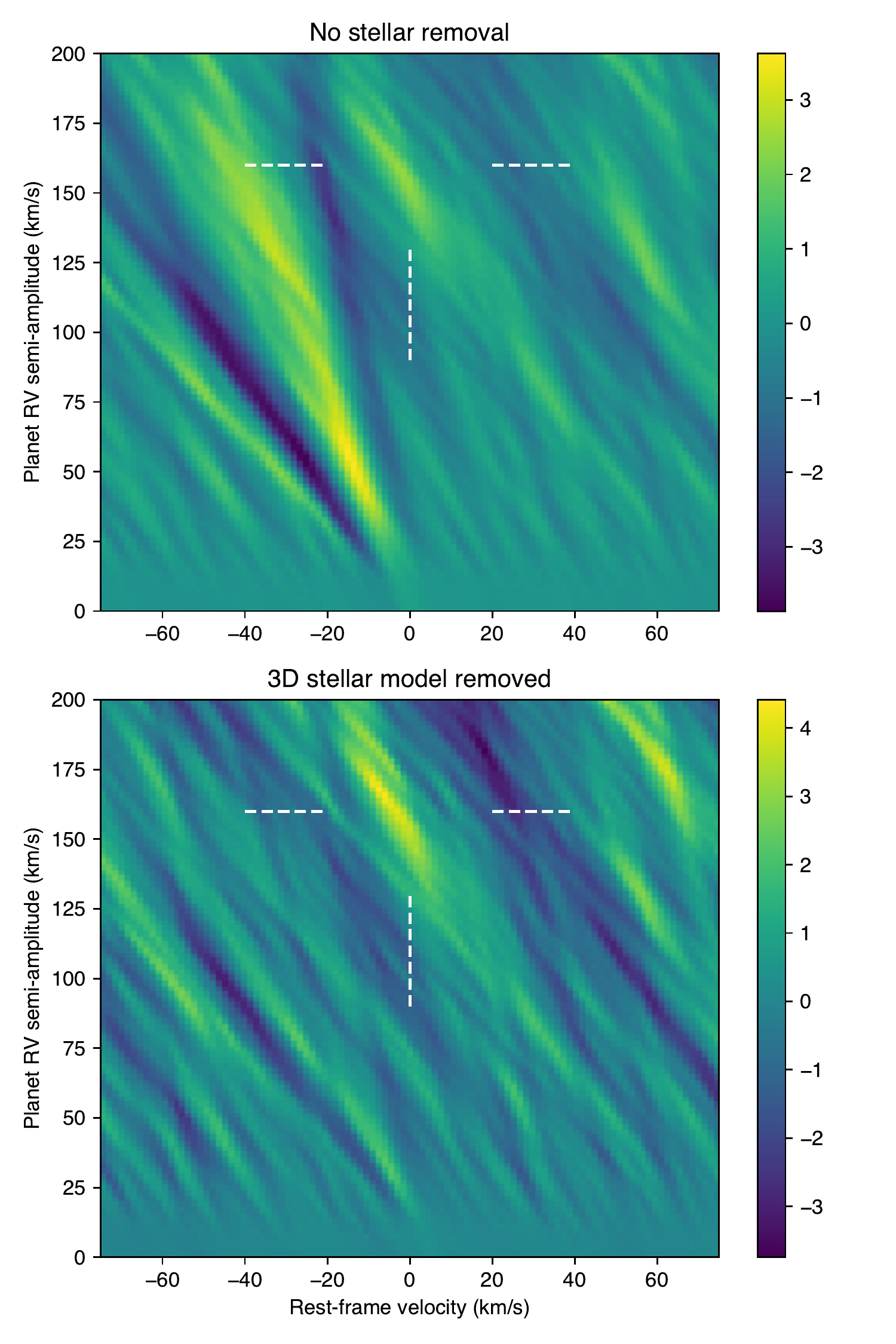}    
	\caption{Total cross correlation S/N (peak divided by standard deviation) obtained from the emission spectra of exoplanet HD~189733~b processed without removing the spectrum of the parent star \citep[top panel, analogous to][]{dekok13} and by removing a spatial and temporal average of the 3D simulation detailed in Section \ref{hd189_tr} (bottom panel). A clear detection of the planet, marked with white dashed lines, appear in the latter case.}
	\label{fig:hd189_emission}
\end{figure}

\section{Application to HD~189733}\label{hd189}

\subsection{Transmission spectra}\label{hd189_tr}
The spectral sequence presented in this section include spectra taken during one transit of exoplanet HD~189733~b and thoroughly described in \citet{2016ApJ...817..106B} and \citet{2019AJ....157..209F}, to which we point the reader for further details. In this context it is worth mentioning that out of the 45 spectra in the sequence, the first 6 are out of transit (before ingress), and the remaining 39 cover ingress, mid-transit, and most of the egress of the planet.

Already \citet{2016ApJ...817..106B} had pointed out that the stellar CO lines were a significant contaminant in transmission spectra, as shown by the left panel of Fig.~\ref{fig:hd189_transmission}. They attempted a correction of the stellar spectrum by modelling the average line profile with micro-turbulence, macro-turbulence, and rotational broadening. This allowed them to pinpoint the CO planetary absorption at the expected planet maximum radial velocity (dashed lines, middle panel) albeit at relatively low significance and with residual signal from the star. A more recent analysis of the same data based on the technique described in this paper \citep{2019AJ....157..209F} achieved a much better correction and resulted in a unique and unambiguous identification of the planetary signal in CO alone (i.e. without the aid of the extra cross-correlation signal from water vapour), as shown in the right panel of Fig.~\ref{fig:hd189_transmission}. This improvement reflected into a refined inference on the rotational rate and wind speed of exoplanet HD~189733~b.

As the process of stellar removal is only briefly explained in \citet{2019AJ....157..209F}, we provide more details in this section. The metallicity of the star HD~189733 is calibrated on a small set of unblended stellar CO lines far from telluric lines. We find that the literature value of [Fe/H] = -0.03 is a good match to the strength of the stellar CO lines, although a solar metallicity spectrum is not visually distinguishable at the noise level of these data. 

The application of the stellar model spectra to transit observation requires a numerical model capable of reproducing the correct geometry of the transit. The model is based closely on the work of \citet{2016ApJ...817..106B} and it involves sub-dividing the stellar disk into a grid of pixels. Each pixel has a different stellar spectrum assigned, based on its value of ($\mu, \phi$) as defined in Section~\ref{sec:3dmodels}, and obtained by bi-linearly interpolating across the temporally-averaged grid of 3D simulations. Furthermore, the spectrum at each pixel is Doppler shifted based on the projected stellar rotational velocity of $v\sin(i) = 3.3$ km s$^{-1}$ \citep{Triaud2009}. Although in this work we assume rigid rotation for the star, the model allows to incorporate differential rotation, which might be necessary once measurements of stellar radial velocity can constrain this quantity \citep[see, e.g.,][]{2016A&A...588A.127C}. Additional radial velocity components, such as a gravitational redshift of 0.68 km s$^{-1}$, the reflex motion due to the orbiting exoplanet (with semi-amplitude 202 m s$^{-1}$) and the combined systemic and barycentric radial velocity are also applied to each time-resolved spectrum.

For observations outside of the planet transit, the total stellar spectrum is given by the sum of the spectra of each individual pixel. This temporally and spatially averaged spectrum will be utilised in Section~\ref{hd189_day} as well to correct dayside observations of the same system. During transit the correct planet-star position needs to be computed in order to determine which pixels of the stellar surface are occulted by the planet disk. As explained in \citet{2016ApJ...817..106B}, we use the known spin-orbit angle of the system \citep[$i_\mathrm{spin}$ = 0.85$^{+0.28}_{-0.32}$ degrees,][]{Triaud2009}, transit impact parameter \citep[$b = 0.6631\pm0.0023$,][]{Agol2010}, and planet orbital phase to solve for the geometry. Once the occulted pixels of the stellar disk are zeroed, the corresponding stellar spectrum at that specific planet phase is computed again by summing all the non-zero pixels. The output stellar model, that is now time-resolved during the planet transit, naturally incorporates the distortions in the line profile due to center-to-limb variations and Rossiter-McLaughlin effect. 
As each 3D RHD simulation spectrum is normalised by the local continuum, the correction of the stellar spectrum is obtained by dividing the sequence of observed spectra through the sequence of modelled spectra, with no further scaling required.

\begin{figure*}
	\centering
		\includegraphics[width=1.\hsize]{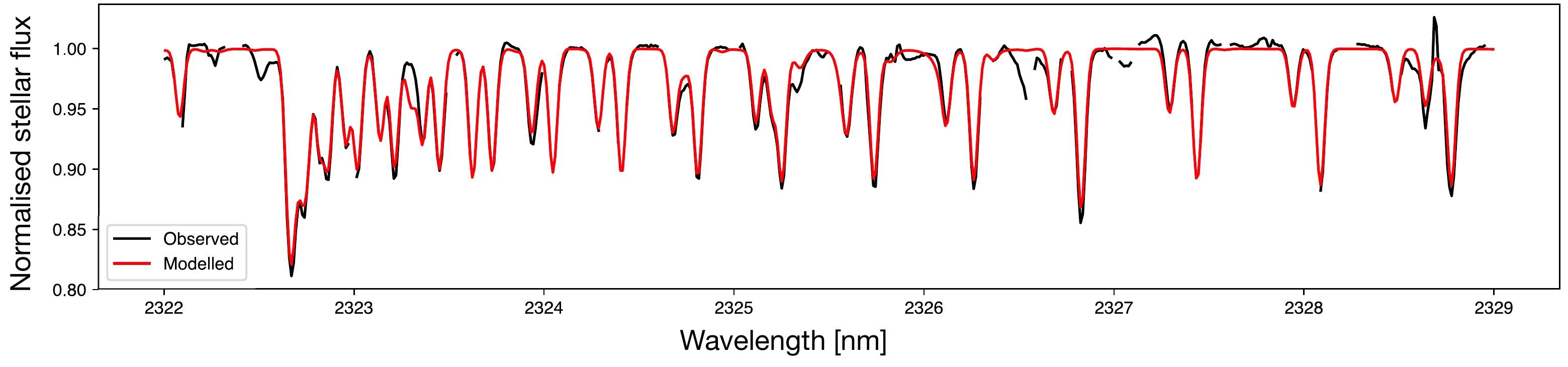}    
	\caption{Average observed spectrum of 51 Pegasi obtained from the entire sequence of CRIRES observations presented in Section~\ref{51peg}, after removal of a telluric model (black line). The corresponding 3D spectrum from Table~\ref{simus}, broadened by $v\sin(i) = 1.5$ km s$^{-1}$ and by the measured CRIRES instrumental profile, is overplotted in red.} 
	\label{fig:51peg_star_spectrum}
\end{figure*}
\subsection{Emission spectra}\label{hd189_day}

110 spectra of HD~189733 with planet 'b' just before secondary eclipse (orbital phases 0.38-0.48) were analysed by \citet{dekok13}. They reported a detection of the planet's thermal spectrum in CO at a S/N=5, but also a significant contamination from stellar CO which they solved by masking those portions of the spectra corresponding to strong stellar absorption. This was possible due to the significant difference between the stellar and planetary radial velocity far from secondary eclipse, however it still produced residual stellar noise due to the sharp edges to the masking filter applied to the data.

We use the 3D stellar spectra of HD 189733 (Table~\ref{simus}) optimised for the data described in Section~\ref{hd189_tr} to correct these observations as well. In this case, since the planet is not transiting, we do not need to accurately model the transit geometry, and a model averaged both spatially and temporally is adopted. This is once again shifted by the systemic and barycentric velocity at the time of observations, and gravitational redshift.

Fig.~\ref{fig:hd189_emission} shows the results of the analysis in the case of uncorrected stellar spectrum (top panel), and after the stellar correction. In the latter case, we recover the signal of exoplanet HD~189733~b in CO at a S/N=4.5, consistent with \citet{dekok13}, and no stellar residual above the S/N = 3. Conversely, without a correction of the stellar spectrum the planet signal would be completely outshone by stellar residuals, preventing its unambiguous identification.

\section{Confirming the atmospheric signature of exoplanet 51 Pegasi b}\label{51peg}

The last set of spectra analysed in this work are described in detail in \citet{brogi13}. It consists of three sequences of spectra taken on three different nights just before, around, and after superior conjunction of the planet. As 51~Pegasi~b is not transiting, its identification in velocity space is complicated by the unknown orbital inclination and poorly known time of inferior conjunction (or any equivalent reference time), which means a large parameter space in the rest-frame velocity ($V_\mathrm{rest}$) versus semi-amplitude of the planet radial velocity ($K_\mathrm{P}$) diagrams such those in Fig.~\ref{hd189_tr} and \ref{hd189_day} is allowed. The presence of stellar contaminating signal can thus significantly complicate or even prevent detection.

In their original study, \citet{brogi13} attempted to scale an empirical solar spectrum to remove a large fraction of the stellar CO spectrum. This approach worked sufficiently well on two of the three nights, i.e. when the planet was far from superior conjunction and its radial velocity departed from the systemic velocity by tens of km s$^{-1}$. However, on the third night (close to superior conjunction) no signal was detected, and no compelling evidence was found to justify the non-detection. We decided to revisit these data and remove the spectral signature of the star 51~Pegasi with application of the models described in this paper.

Fig.~\ref{fig:51peg_star_spectrum} shows the observed spectrum of 51~Pegasi after telluric lines have been removed with the model spectra in output of the ESO Sky Model Calculator, with the fitted values of airmass, PWV, and fitted IP. Overplotted is the 3D RHD simulations used to correct these spectra, which is a clear good match to most of the stellar lines, especially the CO lines clearly identifiable by the band-head near 2.323 $\mu$m. As 51 Pegasi has nearly solar physical parameters, we use solar models for this work, with the exception of stellar metallicity which is set by visual inspection to the value of [Fe/H] = $-0.15$.

\begin{figure}
	\centering
	\includegraphics[width=1.05\hsize]{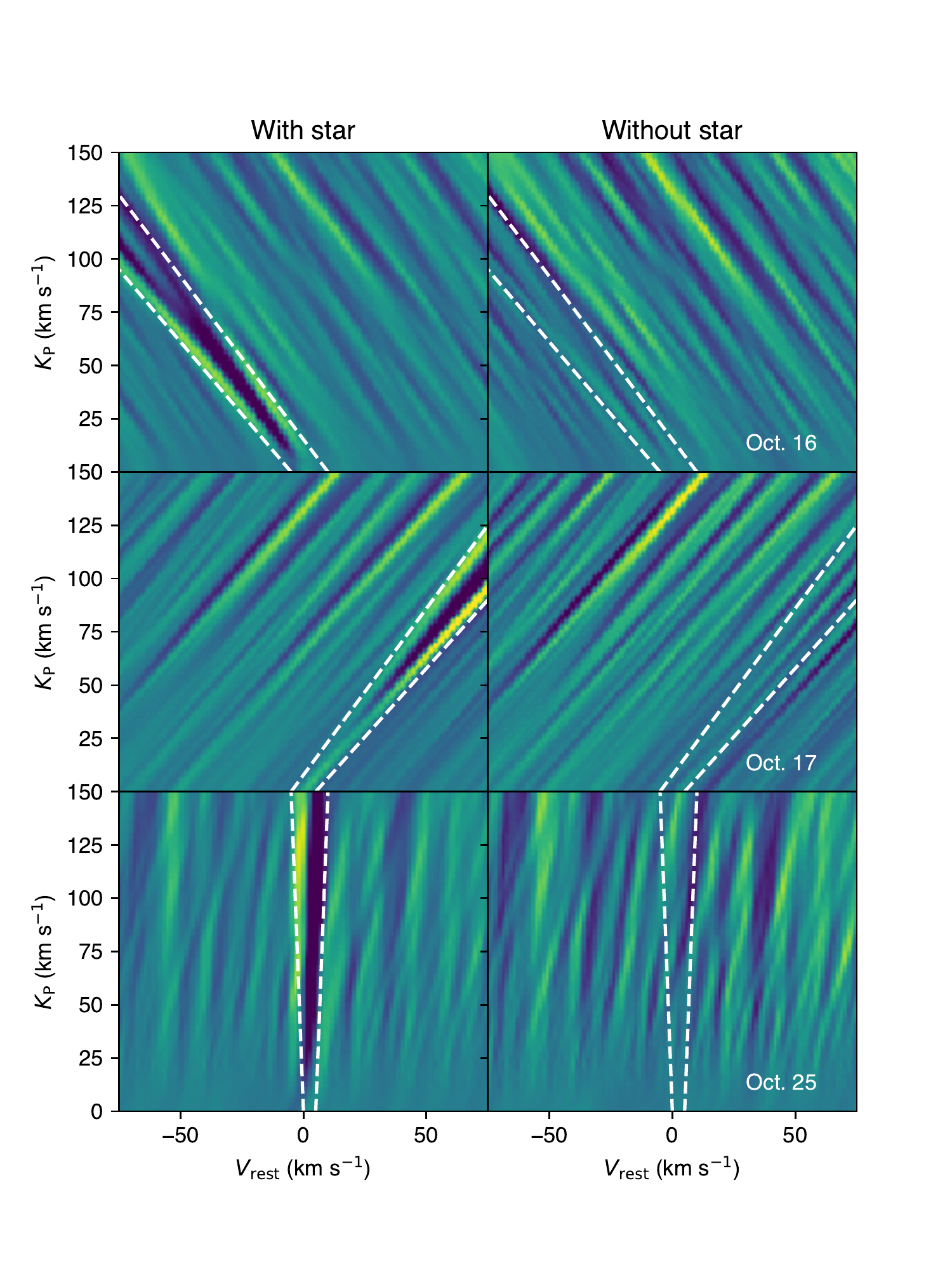}    
	\caption{Total cross correlation signal from single-night observations of 51 Pegasi with VLT/CRIRES, shown as a function of planet rest-frame velocity ($V_\mathrm{rest}$) and maximum orbital radial velocity $K_\mathrm{P}$. The left column shown the analysis without any removal of the stellar spectrum, whereas the right column shows the analysis after removing the stellar spectrum with the tools described in this work. The expected contamination of stellar lines (mostly CO) is marked with dashed white lines. This figure should be compared with Fig.~2 of \citet{brogi13}.} 
	\label{fig:51peg_individual_nights}
\end{figure}

Fig.~\ref{fig:51peg_individual_nights} should be directly compared with Fig.~2 of \citet{brogi13}, and it shows the total cross correlation signal from their best-fitting planetary model, which contains both CO and H$_2$O, when the stellar CO lines are not corrected (left panel) and when they are removed with the analysis described in this paper (right panel). Similarly to their work, the stellar correction manages to bring the residual stellar cross correlation below the level of the noise, while the planet signal, barely visible on individual nights as a bright stripe of positive correlation around $V_\mathrm{rest} = 0$ and $K_\mathrm{P} = 132$ km s$^{-1}$, is preserved in the process. The overall performances of the stellar removal appear superior to the analysis in \citet{brogi13}, and this is further demonstrated when the individual observing nights are co-added.

\begin{figure}
	\centering
	\includegraphics[width=1.05\hsize]{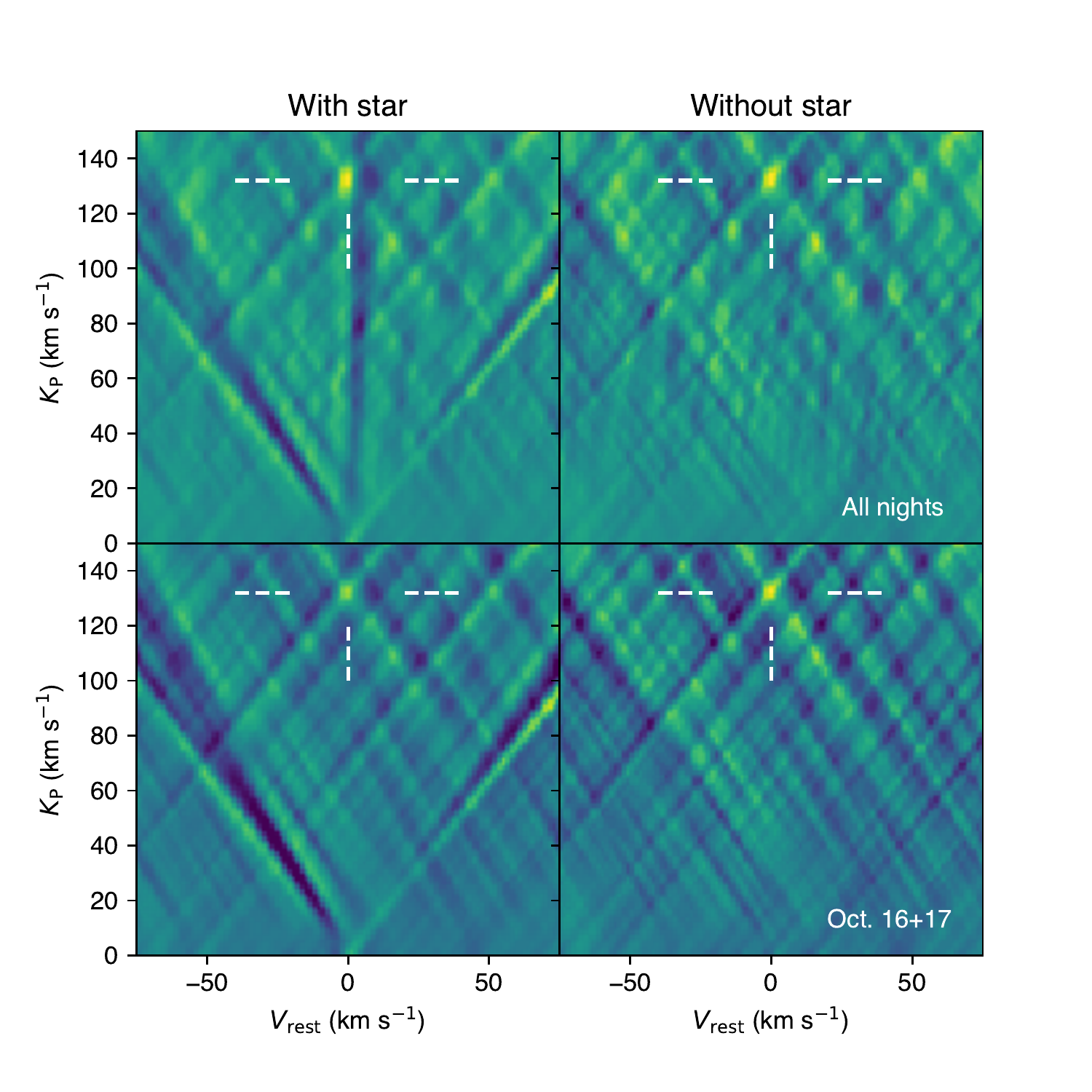}    
	\caption{Total cross correlation signal obtained by co-adding the 3 nights of data of 51 Pegasi (top row), and the nights of October 16$^\mathrm{th}$ and 17$^\mathrm{th}$ only (bottom row). The quantities on the axes are analogous to Fig.~\ref{fig:51peg_individual_nights}. The known position of exoplanet 51 Peg b is marked with white dashed lines. Although the exoplanet cross-correlation signal is detected even without removing the stellar spectrum, it is confused in the strong stellar residuals (left panel). Conversely, after removal of the stellar spectrum the planetary signal is the only unambiguous signature detected, and all three nights co-add constructively as expected.}
	\label{fig:51peg_comb_CCF}
\end{figure}

Fig.~\ref{fig:51peg_comb_CCF} shows the combined cross correlation signal from the 3 observing nights (top row) and from the nights of 16 and 17 October only (bottomo row). The stellar residuals make a striking difference in the co-added signal. Although a clear positive deviation in the total cross correlation signal appears at the known position of the planet regardless of whether the star is removed or not, in the latter case its planetary nature cannot be unambiguously determined. The presence of strong stellar residuals near the planet position on the night of October 25, which is a consequence of the similar planet and stellar radial velocities near superior conjunction, makes the planet detection doubtful.

Conversely, when the stellar spectrum is removed, the planet 51 Pegasi b remains the only unambiguous source detected. Remarkably, the planet is now detected in all three nights of observation, with a noticeable gain in S/N (about 10\%) when adding the night of October 25. This is a clear improvement from the non-detection originally reported by \citet{brogi13} on this night, which allows us to confirm the measurement of molecular absorption in $K$-band spectra of this planet for the first time.

\subsection{Bisectors}

In Section~\ref{sec:stellarvar}, we showed the convection plays a crucial role in the formation of spectral lines and deeply influences the shape, shift, and asymmetries of lines in late type stars. One way to detect the asymmetries in the line is the bisector, defined as the locus of the midpoints of the spectral line. A symmetric profile has a straight vertical bisector and is the consequence of a complete absence of velocity-brightness correlations (i.e., the case of 1D hydrostatic models), while bisectors with C-shape are formed mostly in the upflows (granules) or reverse C-shape bisectors are generally formed in downflows \citep{1981A&A....96..345D}. In the literature, bisectors of several kinds of stars has already reavealed the presence of asymmetries and wavelength shifts caused by the granulation \citep[e.g., ][]{2008A&A...492..841R,2009ApJ...697.1032G}. 

\begin{figure}
   \centering
    \includegraphics[width=1.0\hsize]{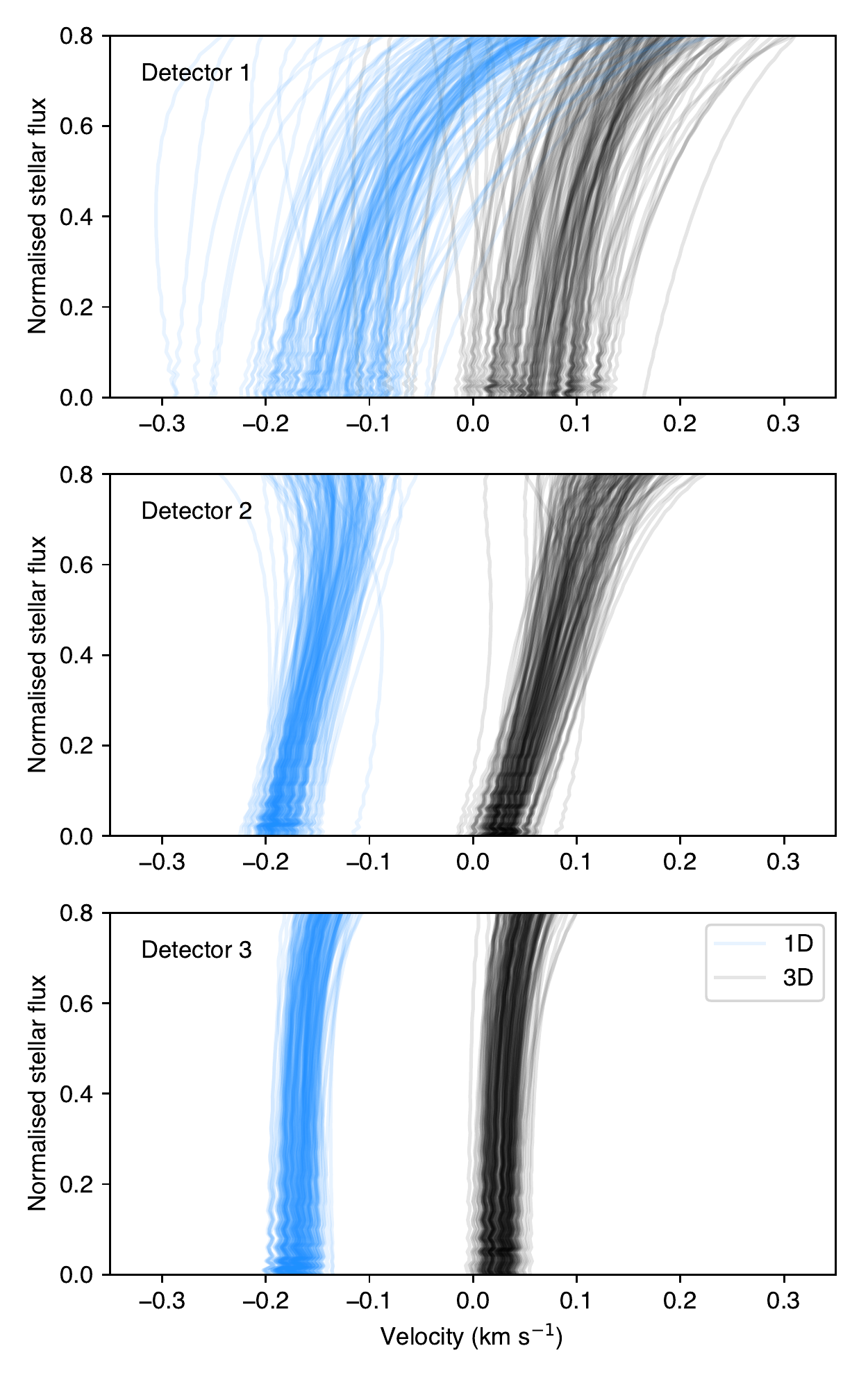}
      \caption{Bisectors obtained from the cross cross correlation function of the observed spectrum of 51 Pegasi (shown in Fig. \ref{fig:51peg_star_spectrum}) with 1D hydrostatic models (light blue) and 3D RHD simulation (black). From top to bottom we show the different CRIRES detectors, except detector number 4 which is known to suffer from odd-even effect and will likely have an asymmetric instrument profile. Each bisector results from one of the 166 exposures observed on the night of October 16, 2010.}
        \label{bisectors}
   \end{figure}

In this section, we aim to check weather our use of numerical stellar models is correct. Initially we cross correlate the stellar spectrum 51~Pegasi, resolved temporally, with both the 3D RHD simulation from Table~\ref{simus} and the 1D hydrostatic model from MARCS package \citep{2008A&A...486..951G} with the same stellar parameters as of 3D simulation. We run the cross correlation at $R=10^6$, i.e. 10$\times$ the resolution of CRIRES, resulting into cross correlation functions sampled with a step size of 0.3 km s$^{-1}$. \\
Subsequently, we compute the bisectors for the resulting cross correlated functions as reported in Fig.~\ref{bisectors} (light blue for 1D and black for 3D). A straight and zero-centered bisector would indicate that the model we used to correct the stellar signal is perfectly matching the stellar properties of the star. First point of concern is the shape of bisectors: blue and dark curves are similar across different detectors. This indicates that the CO lines are comparable, in term of shape, in 1D and 3D. Second point, more important, is the bisector shifts. The use of 3D simulation returns systematically values between 0 and 50 m s$^{-1}$. This velocity shift with respect to zero might be due to the gravitational redshift \citep{2003A&A...401.1185L}. In this work we have assumed a gravitational redshift of 650 m s$^{-1}$ for 51 Pegasi, and added this extra radial-velocity component when comparing the synthetic spectra to observations. However, the exact values for 51~Pegasi is unknown even though, for late-type dwarfs, (log$g\approx4.5$), the gravitational shifts ranges between 0.7 and 0.8 km/s and it dramatically decreases with surface gravity down to 0.02-0.03 km/s for K giant stars with log$g\approx1.5$ \citep{2013A&A...550A.103A}. Another possibility for the residual shift of bisectors obtained from 3D spectra is the similar uncertainty in the systemic velocity, which is -33.218$^{+0.076}_{-0.077}$ km s$^{-1}$ according to \citet{birkby2017}. In terms of our measurements, gravitational redshift and systemic velocity could technically be separated because the former only influences the star, whereas the latter influences both stellar and planetary radial velocity. However, this would require a precision less than 100 m/s in the planetary radial velocity, which is nearly one order of magnitude below the current precision \citep[just below 1 km/s for the best planet detections, see e.g.][]{2019AJ....157..209F}. Therefore, for the purposes of this study, one can consider systemic velocity and gravitational redshift indistinguishable and producing just a net shift of the stellar lines.

Whatever the case, cross correlation functions obtained with 1D model show a residual blueshift between -150 and -200 m s$^{-1}$, indicating that 3D model are substantially more adequate to accurately reproduce the photospheric velocity field of the star. 

\section{Conclusions}

We used 3D RHD stellar convection simulations to provide synthetic stellar spectra resolved both spatially and temporally, and we coupled them with an analytical model of a transiting exoplanet accounting for the variable portion of the occulted stellar disk.
We applied the method to VLT/CRIRES observations and removed the spectrum of two bright exoplanets host stars: the early K-dwarf HD~189733 (covering transmission and emission spectroscopy of its transiting planet), and the solar-type 51~Pegasi (covering emission spectroscopy of its non-transiting planet).

Removing the stellar spectrum with our method led to a significantly improved detection of the exoplanet atmosphere. We show that our modelling is superior to a simple parametrisation of the stellar line profile or to the use of one-dimensional stellar models. This is due to the fact that 3D RHD simulations describe correctly the stellar properties and dynamics. 3D spectra match the observed asymmetry and/or blue-shifted in spectral lines, and intrinsically, they also model center-to-limb variation and Rossiter-McLaughlin effect potentially altering the interpretation of exoplanet transmission spectra. In the case of 51~Pegasi, we succeeded at confirming a previous tentative detection of the planet's K-band spectrum due to the improved modelling of the stellar residuals.

To further establish the synergy between stellar modelling and exoplanet observations it will also be fundamental to extend this work to bluer wavelengths covering the helium triplet at 10830 \AA \citep{oklopcic2018, 2018Sci...362.1388N, 2018Sci...362.1384A}, and the alkali lines in the optical \citep[Na and K doublet, see e.g.][]{2015A&A...577A..62W, casasayas-barris2017}. However, this is also potentially complicated by the necessity of treating non-LTE effects \citep[e.g.,][]{2018A&A...615A.139A} and by the general limitation in the knowledge of atomic and molecular linelists \citep{2018arXiv181108041J}.\\
Another important aspect is also the extension of the method described in this work to M-dwarf stars that are increasing in number of detections due to their abundance in the solar neighbourhood and the favourable planet/star contrast \citep{anglada2016, gillon2017}. Preliminary 3D global (i.e., including the whole convective surface layers within the numerical box, star-in-a-box configuration) RHD simulation have been achieved in \cite{2013MSAIS..24..128A} in presence of rotation, essential to resolve the cloud surface distribution and to induced potential variability.
However, M dwarfs are magnetically active with possibly local suppression of convection by magnetic field lines emerging at the surface. Thus, global Magneto-RHD simulations are needed to identify the mixing length to use in classical 1D models to compensate for this suppression \citep{2013MSAIS..24..128A}.

In conclusion, three-dimensional RHD simulations are now established as realistic descriptions for the convective photospheres of various classes of stars. The good and time-dependent representation of the background stellar disk is a natural and necessary step forward toward a better understanding of stellar properties. We have now proven that, in the context of exoplanet science, 3D RHD simulations are also extremely useful for a detailed and quantitative analysis of the atmospheric signatures of transiting and non-transiting planets.

\bibliographystyle{aa}
\bibliography{biblio.bib}

\begin{thebibliography}{91}
\expandafter\ifx\csname natexlab\endcsname\relax\def\natexlab#1{#1}\fi

\bibitem[{{Agol} {et~al.}(2010){Agol}, {Cowan}, {Knutson}, {Deming}, {Steffen},
  {Henry}, \& {Charbonneau}}]{Agol2010}
{Agol}, E., {Cowan}, N.~B., {Knutson}, H.~A., {et~al.} 2010, \apj, 721, 1861

\bibitem[{{Aigrain} {et~al.}(2004){Aigrain}, {Favata}, \&
  {Gilmore}}]{2004A&A...414.1139A}
{Aigrain}, S., {Favata}, F., \& {Gilmore}, G. 2004, \aap, 414, 1139

\bibitem[{{Allard} {et~al.}(2013){Allard}, {Homeier}, {Freytag},
  {Schaffenberger}, {}, \& {Rajpurohit}}]{2013MSAIS..24..128A}
{Allard}, F., {Homeier}, D., {Freytag}, B., {et~al.} 2013, Memorie della
  Societa Astronomica Italiana Supplementi, 24, 128

\bibitem[{{Allart} {et~al.}(2018){Allart}, {Bourrier}, {Lovis}, {Ehrenreich},
  {Spake}, {Wyttenbach}, {Pino}, {Pepe}, {Sing}, \& {Lecavelier des
  Etangs}}]{2018Sci...362.1384A}
{Allart}, R., {Bourrier}, V., {Lovis}, C., {et~al.} 2018, Science, 362, 1384

\bibitem[{{Allende Prieto} {et~al.}(2013){Allende Prieto}, {Koesterke},
  {Ludwig}, {Freytag}, \& {Caffau}}]{2013A&A...550A.103A}
{Allende Prieto}, C., {Koesterke}, L., {Ludwig}, H.-G., {Freytag}, B., \&
  {Caffau}, E. 2013, \aap, 550, A103

\bibitem[{{Amarsi} {et~al.}(2018){Amarsi}, {Nordlander}, {Barklem}, {Asplund},
  {Collet}, \& {Lind}}]{2018A&A...615A.139A}
{Amarsi}, A.~M., {Nordlander}, T., {Barklem}, P.~S., {et~al.} 2018, \aap, 615,
  A139

\bibitem[{{Anglada-Escud{\'e}} {et~al.}(2016){Anglada-Escud{\'e}}, {Amado},
  {Barnes}, {Berdi{\~n}as}, {Butler}, {Coleman}, {de La Cueva}, {Dreizler},
  {Endl}, {Giesers}, {Jeffers}, {Jenkins}, {Jones}, {Kiraga}, {K{\"u}rster},
  {L{\'o}pez-Gonz{\'a}lez}, {Marvin}, {Morales}, {Morin}, {Nelson}, {Ortiz},
  {Ofir}, {Paardekooper}, {Reiners}, {Rodr{\'\i}guez},
  {Rodr{\'\i}guez-L{\'o}pez}, {Sarmiento}, {Strachan}, {Tsapras}, {Tuomi}, \&
  {Zechmeister}}]{anglada2016}
{Anglada-Escud{\'e}}, G., {Amado}, P.~J., {Barnes}, J., {et~al.} 2016, \nat,
  536, 437

\bibitem[{{Artigau} {et~al.}(2014){Artigau}, {Kouach}, {Donati}, {Doyon},
  {Delfosse}, {Baratchart}, {Lacombe}, {Moutou}, {Rabou}, {Par{\`e}s},
  {Micheau}, {Thibault}, {Reshetov}, {Dubois}, {Hernandez}, {Vall{\'e}e},
  {Wang}, {Dolon}, {Pepe}, {Bouchy}, {Striebig}, {H{\'e}nault}, {Loop},
  {Saddlemyer}, {Barrick}, {Vermeulen}, {Dupieux}, {H{\'e}brard}, {Boisse},
  {Martioli}, {Alencar}, {do Nascimento}, \& {Figueira}}]{spirou2014}
{Artigau}, {\'E}., {Kouach}, D., {Donati}, J.-F., {et~al.} 2014, in \procspie,
  Vol. 9147, 914715

\bibitem[{{Asplund}(2000)}]{2000A&A...359..755A}
{Asplund}, M. 2000, \aap, 359, 755

\bibitem[{{Asplund} {et~al.}(2005){Asplund}, {Grevesse}, \&
  {Sauval}}]{2005ASPC..336...25A}
{Asplund}, M., {Grevesse}, N., \& {Sauval}, A.~J. 2005, in Astronomical Society
  of the Pacific Conference Series, Vol. 336, Cosmic Abundances as Records of
  Stellar Evolution and Nucleosynthesis, ed. T.~G. {Barnes}, III \& F.~N.
  {Bash}, 25

\bibitem[{{Asplund} {et~al.}(2009{\natexlab{a}}){Asplund}, {Grevesse},
  {Sauval}, \& {Scott}}]{2009ARA&A..47..481A}
{Asplund}, M., {Grevesse}, N., {Sauval}, A.~J., \& {Scott}, P.
  2009{\natexlab{a}}, \araa, 47, 481

\bibitem[{{Asplund} {et~al.}(2009{\natexlab{b}}){Asplund}, {Grevesse},
  {Sauval}, \& {Scott}}]{asplund09}
{Asplund}, M., {Grevesse}, N., {Sauval}, A.~J., \& {Scott}, P.
  2009{\natexlab{b}}, \araa, 47, 481

\bibitem[{{Asplund} {et~al.}(2000{\natexlab{a}}){Asplund}, {Ludwig},
  {Nordlund}, \& {Stein}}]{2000A&A...359..669A}
{Asplund}, M., {Ludwig}, H., {Nordlund}, {\AA}., \& {Stein}, R.~F.
  2000{\natexlab{a}}, \aap, 359, 669

\bibitem[{{Asplund} {et~al.}(2000{\natexlab{b}}){Asplund}, {Nordlund},
  {Trampedach}, \& {Stein}}]{2000A&A...359..743A}
{Asplund}, M., {Nordlund}, {\AA}., {Trampedach}, R., \& {Stein}, R.~F.
  2000{\natexlab{b}}, \aap, 359, 743

\bibitem[{{Barclay} {et~al.}(2018){Barclay}, {Pepper}, \&
  {Quintana}}]{2018ApJS..239....2B}
{Barclay}, T., {Pepper}, J., \& {Quintana}, E.~V. 2018, \apjs, 239, 2

\bibitem[{{Beeck} {et~al.}(2013{\natexlab{a}}){Beeck}, {Cameron}, {Reiners}, \&
  {Sch{\"u}ssler}}]{2013A&A...558A..48B}
{Beeck}, B., {Cameron}, R.~H., {Reiners}, A., \& {Sch{\"u}ssler}, M.
  2013{\natexlab{a}}, \aap, 558, A48

\bibitem[{{Beeck} {et~al.}(2013{\natexlab{b}}){Beeck}, {Cameron}, {Reiners}, \&
  {Sch{\"u}ssler}}]{2013A&A...558A..49B}
{Beeck}, B., {Cameron}, R.~H., {Reiners}, A., \& {Sch{\"u}ssler}, M.
  2013{\natexlab{b}}, \aap, 558, A49

\bibitem[{{Bigot} {et~al.}(2011){Bigot}, {Mourard}, {Berio}, {Th{\'e}venin},
  {Ligi}, {Tallon-Bosc}, {Chesneau}, {Delaa}, {Nardetto}, {Perraut}, {Stee},
  {Boyajian}, {Morel}, {Pichon}, {Kervella}, {Schmider}, {McAlister}, {ten
  Brummelaar}, {Ridgway}, {Sturmann}, {Sturmann}, {Turner}, {Farrington}, \&
  {Goldfinger}}]{2011A&A...534L...3B}
{Bigot}, L., {Mourard}, D., {Berio}, P., {et~al.} 2011, \aap, 534, L3

\bibitem[{{Bigot} \& {Th{\'e}venin}(2008)}]{2008sf2a.conf....3B}
{Bigot}, L. \& {Th{\'e}venin}, F. 2008, in Proceedings of the Annual meeting of
  the French Society of Astronomy and Astrophysics, ed. C.~{Charbonnel},
  F.~{Combes}, \& R.~{Samadi}, 3

\bibitem[{{Birkby}(2018)}]{birkby_review}
{Birkby}, J.~L. 2018, Handbook of Exoplanets, Springer Nature, 16

\bibitem[{{Birkby} {et~al.}(2017){Birkby}, {de Kok}, {Brogi}, {Schwarz}, \&
  {Snellen}}]{birkby2017}
{Birkby}, J.~L., {de Kok}, R.~J., {Brogi}, M., {Schwarz}, H., \& {Snellen},
  I.~A.~G. 2017, \aj, 153, 138

\bibitem[{{B{\"o}hm-Vitense}(1958)}]{1958ZA.....46..108B}
{B{\"o}hm-Vitense}, E. 1958, \zap, 46, 108

\bibitem[{{Bonifacio} {et~al.}(2017){Bonifacio}, {Caffau}, {Ludwig}, {Steffen},
  {Castelli}, {Gallagher}, {Prakapavi{\v c}ius}, {Ku{\v c}inskas}, {Cayrel},
  {Freytag}, {Plez}, \& {Homeier}}]{2017MmSAI..88...90B}
{Bonifacio}, P., {Caffau}, E., {Ludwig}, H.-G., {et~al.} 2017, \memsai, 88, 90

\bibitem[{{Brogi} {et~al.}(2016){Brogi}, {de Kok}, {Albrecht}, {Snellen},
  {Birkby}, \& {Schwarz}}]{2016ApJ...817..106B}
{Brogi}, M., {de Kok}, R.~J., {Albrecht}, S., {et~al.} 2016, \apj, 817, 106

\bibitem[{{Brogi} {et~al.}(2017){Brogi}, {Line}, {Bean}, {D{\'e}sert}, \&
  {Schwarz}}]{2017ApJ...839L...2B}
{Brogi}, M., {Line}, M., {Bean}, J., {D{\'e}sert}, J.-M., \& {Schwarz}, H.
  2017, \apjl, 839, L2

\bibitem[{{Brogi} \& {Line}(2019)}]{2019AJ....157..114B}
{Brogi}, M. \& {Line}, M.~R. 2019, \aj, 157, 114

\bibitem[{{Brogi} {et~al.}(2012){Brogi}, {Snellen}, {de Kok}, {Albrecht},
  {Birkby}, \& {de Mooij}}]{2012Natur.486..502B}
{Brogi}, M., {Snellen}, I.~A.~G., {de Kok}, R.~J., {et~al.} 2012, \nat, 486,
  502

\bibitem[{{Brogi} {et~al.}(2013){Brogi}, {Snellen}, {de Kok}, {Albrecht},
  {Birkby}, \& {de Mooij}}]{brogi13}
{Brogi}, M., {Snellen}, I.~A.~G., {de Kok}, R.~J., {et~al.} 2013, \apj, 767, 27

\bibitem[{{Caffau} {et~al.}(2011){Caffau}, {Ludwig}, {Steffen}, {Freytag}, \&
  {Bonifacio}}]{2011SoPh..268..255C}
{Caffau}, E., {Ludwig}, H.-G., {Steffen}, M., {Freytag}, B., \& {Bonifacio}, P.
  2011, \solphys, 268, 255

\bibitem[{{Casasayas-Barris} {et~al.}(2017){Casasayas-Barris}, {Palle},
  {Nowak}, {Yan}, {Nortmann}, \& {Murgas}}]{casasayas-barris2017}
{Casasayas-Barris}, N., {Palle}, E., {Nowak}, G., {et~al.} 2017, \aap, 608,
  A135

\bibitem[{{Cegla} {et~al.}(2016){Cegla}, {Lovis}, {Bourrier}, {Beeck},
  {Watson}, \& {Pepe}}]{2016A&A...588A.127C}
{Cegla}, H.~M., {Lovis}, C., {Bourrier}, V., {et~al.} 2016, \aap, 588, A127

\bibitem[{{Cegla} {et~al.}(2018){Cegla}, {Watson}, {Shelyag}, {Chaplin},
  {Davies}, {Mathioudakis}, {Palumbo}, {Saar}, \&
  {Haywood}}]{2018ApJ...866...55C}
{Cegla}, H.~M., {Watson}, C.~A., {Shelyag}, S., {et~al.} 2018, \apj, 866, 55

\bibitem[{{Chiavassa} {et~al.}(2012){Chiavassa}, {Bigot}, {Kervella}, {Matter},
  {Lopez}, {Collet}, {Magic}, \& {Asplund}}]{2012A&A...540A...5C}
{Chiavassa}, A., {Bigot}, L., {Kervella}, P., {et~al.} 2012, \aap, 540, A5

\bibitem[{{Chiavassa} {et~al.}(2011){Chiavassa}, {Bigot}, {Th{\'e}venin},
  {Collet}, {Jasniewicz}, {Magic}, \& {Asplund}}]{2011JPhCS.328a2012C}
{Chiavassa}, A., {Bigot}, L., {Th{\'e}venin}, F., {et~al.} 2011, Journal of
  Physics Conference Series, 328, 012012

\bibitem[{{Chiavassa} {et~al.}(2017){Chiavassa}, {Caldas}, {Selsis}, {Leconte},
  {Von Paris}, {Bord{\'e}}, {Magic}, {Collet}, \&
  {Asplund}}]{2017A&A...597A..94C}
{Chiavassa}, A., {Caldas}, A., {Selsis}, F., {et~al.} 2017, \aap, 597, A94

\bibitem[{{Chiavassa} {et~al.}(2018){Chiavassa}, {Casagrande}, {Collet},
  {Magic}, {Bigot}, {Th{\'e}venin}, \& {Asplund}}]{2018A&A...611A..11C}
{Chiavassa}, A., {Casagrande}, L., {Collet}, R., {et~al.} 2018, \aap, 611, A11

\bibitem[{{Chiavassa} {et~al.}(2010){Chiavassa}, {Collet}, {Casagrande}, \&
  {Asplund}}]{2010A&A...524A..93C}
{Chiavassa}, A., {Collet}, R., {Casagrande}, L., \& {Asplund}, M. 2010, \aap,
  524, A93

\bibitem[{{Chiavassa} {et~al.}(2015){Chiavassa}, {Pere}, {Faurobert}, {Ricort},
  {Tanga}, {Magic}, {Collet}, \& {Asplund}}]{2015A&A...576A..13C}
{Chiavassa}, A., {Pere}, C., {Faurobert}, M., {et~al.} 2015, \aap, 576, A13

\bibitem[{{Chiavassa} {et~al.}(2009){Chiavassa}, {Plez}, {Josselin}, \&
  {Freytag}}]{2009A&A...506.1351C}
{Chiavassa}, A., {Plez}, B., {Josselin}, E., \& {Freytag}, B. 2009, \aap, 506,
  1351

\bibitem[{{Collet} {et~al.}(2011){Collet}, {Hayek}, {Asplund}, {Nordlund},
  {Trampedach}, \& {Gudiksen}}]{2011A&A...528A..32C}
{Collet}, R., {Hayek}, W., {Asplund}, M., {et~al.} 2011, \aap, 528, A32

\bibitem[{{Creevey} {et~al.}(2012){Creevey}, {Th{\'e}venin}, {Boyajian},
  {Kervella}, {Chiavassa}, {Bigot}, {M{\'e}rand}, {Heiter}, {Morel}, {Pichon},
  {Mc Alister}, {ten Brummelaar}, {Collet}, {van Belle}, {Coud{\'e} du
  Foresto}, {Farrington}, {Goldfinger}, {Sturmann}, {Sturmann}, \&
  {Turner}}]{2012A&A...545A..17C}
{Creevey}, O.~L., {Th{\'e}venin}, F., {Boyajian}, T.~S., {et~al.} 2012, \aap,
  545, A17

\bibitem[{{de Kok} {et~al.}(2013){de Kok}, {Brogi}, {Snellen}, {Birkby},
  {Albrecht}, \& {de Mooij}}]{dekok13}
{de Kok}, R.~J., {Brogi}, M., {Snellen}, I.~A.~G., {et~al.} 2013, \aap, 554,
  A82

\bibitem[{{Dravins}(1982)}]{1982ARA&A..20...61D}
{Dravins}, D. 1982, \araa, 20, 61

\bibitem[{{Dravins}(1987)}]{dravins87}
{Dravins}, D. 1987, \aap, 172, 211

\bibitem[{{Dravins} {et~al.}(2018){Dravins}, {Gustavsson}, \&
  {Ludwig}}]{2018A&A...616A.144D}
{Dravins}, D., {Gustavsson}, M., \& {Ludwig}, H.-G. 2018, \aap, 616, A144

\bibitem[{{Dravins} {et~al.}(1981){Dravins}, {Lindegren}, \&
  {Nordlund}}]{1981A&A....96..345D}
{Dravins}, D., {Lindegren}, L., \& {Nordlund}, A. 1981, \aap, 96, 345

\bibitem[{{Dravins} {et~al.}(2017){Dravins}, {Ludwig}, {Dahl{\'e}n}, \&
  {Pazira}}]{2017A&A...605A..90D}
{Dravins}, D., {Ludwig}, H.-G., {Dahl{\'e}n}, E., \& {Pazira}, H. 2017, \aap,
  605, A90

\bibitem[{{Eistrup} {et~al.}(2018){Eistrup}, {Walsh}, \& {van
  Dishoeck}}]{2018A&A...613A..14E}
{Eistrup}, C., {Walsh}, C., \& {van Dishoeck}, E.~F. 2018, \aap, 613, A14

\bibitem[{{Flowers} {et~al.}(2019){Flowers}, {Brogi}, {Rauscher}, {Kempton}, \&
  {Chiavassa}}]{2019AJ....157..209F}
{Flowers}, E., {Brogi}, M., {Rauscher}, E., {Kempton}, E. M.~R., \&
  {Chiavassa}, A. 2019, \aj, 157, 209

\bibitem[{{Follert} {et~al.}(2014){Follert}, {Dorn}, {Oliva}, {Lizon},
  {Hatzes}, {Piskunov}, {Reiners}, {Seemann}, {Stempels}, {Heiter}, {Marquart},
  {Lockhart}, {Anglada-Escude}, {L{\"o}winger}, {Baade}, {Grunhut}, {Bristow},
  {Klein}, {Jung}, {Ives}, {Kerber}, {Pozna}, {Paufique}, {Kaeufl}, {Origlia},
  {Valenti}, {Gojak}, {Hilker}, {Pasquini}, {Smette}, \&
  {Smoker}}]{criresplus2014}
{Follert}, R., {Dorn}, R.~J., {Oliva}, E., {et~al.} 2014, in \procspie, Vol.
  9147, 914719

\bibitem[{{Foreman-Mackey} {et~al.}(2013){Foreman-Mackey}, {Hogg}, {Lang}, \&
  {Goodman}}]{foreman13}
{Foreman-Mackey}, D., {Hogg}, D.~W., {Lang}, D., \& {Goodman}, J. 2013, \pasp,
  125, 306

\bibitem[{{Frohlich} {et~al.}(1997){Frohlich}, {Andersen}, {Appourchaux},
  {Berthomieu}, {Crommelynck}, {Domingo}, {Fichot}, {Finsterle}, {Gomez},
  {Gough}, {Jimenez}, {Leifsen}, {Lombaerts}, {Pap}, {Provost}, {Cortes},
  {Romero}, {Roth}, {Sekii}, {Telljohann}, {Toutain}, \&
  {Wehrli}}]{1997SoPh..170....1F}
{Frohlich}, C., {Andersen}, B.~N., {Appourchaux}, T., {et~al.} 1997, \solphys,
  170, 1

\bibitem[{{Gandolfi} {et~al.}(2018){Gandolfi}, {Barrag{\'a}n}, {Livingston},
  {Fridlund}, {Justesen}, {Redfield}, {Fossati}, {Mathur}, {Grziwa}, {Cabrera},
  {Garc{\'\i}a}, {Persson}, {Van Eylen}, {Hatzes}, {Hidalgo}, {Albrecht},
  {Bugnet}, {Cochran}, {Csizmadia}, {Deeg}, {Eigm{\"u}ller}, {Endl}, {Erikson},
  {Esposito}, {Guenther}, {Korth}, {Luque}, {Monta{\~n}es Rodr{\'\i}guez},
  {Nespral}, {Nowak}, {P{\"a}tzold}, \& {Prieto-Arranz}}]{pimensae}
{Gandolfi}, D., {Barrag{\'a}n}, O., {Livingston}, J.~H., {et~al.} 2018, \aap,
  619, L10

\bibitem[{{Gillon} {et~al.}(2017){Gillon}, {Triaud}, {Demory}, {Jehin}, {Agol},
  {Deck}, {Lederer}, {de Wit}, {Burdanov}, {Ingalls}, {Bolmont}, {Leconte},
  {Raymond}, {Selsis}, {Turbet}, {Barkaoui}, {Burgasser}, {Burleigh}, {Carey},
  {Chaushev}, {Copperwheat}, {Delrez}, {Fernand es}, {Holdsworth}, {Kotze},
  {Van Grootel}, {Almleaky}, {Benkhaldoun}, {Magain}, \& {Queloz}}]{gillon2017}
{Gillon}, M., {Triaud}, A. H.~M.~J., {Demory}, B.-O., {et~al.} 2017, \nat, 542,
  456

\bibitem[{{Gray}(2005)}]{2005oasp.book.....G}
{Gray}, D.~F. 2005, {The Observation and Analysis of Stellar Photospheres}
  (Cambridge University Press)

\bibitem[{{Gray}(2009)}]{2009ApJ...697.1032G}
{Gray}, D.~F. 2009, \apj, 697, 1032

\bibitem[{{Gustafsson} {et~al.}(2008){Gustafsson}, {Edvardsson}, {Eriksson},
  {J{\o}rgensen}, {Nordlund}, \& {Plez}}]{2008A&A...486..951G}
{Gustafsson}, B., {Edvardsson}, B., {Eriksson}, K., {et~al.} 2008, \aap, 486,
  951

\bibitem[{{Hayek} {et~al.}(2010){Hayek}, {Asplund}, {Carlsson}, {Trampedach},
  {Collet}, {Gudiksen}, {Hansteen}, \& {Leenaarts}}]{2010A&A...517A..49H}
{Hayek}, W., {Asplund}, M., {Carlsson}, M., {et~al.} 2010, \aap, 517, A49

\bibitem[{{Husser} {et~al.}(2013){Husser}, {Wende-von Berg}, {Dreizler},
  {Homeier}, {Reiners}, {Barman}, \& {Hauschildt}}]{2013A&A...553A...6H}
{Husser}, T.~O., {Wende-von Berg}, S., {Dreizler}, S., {et~al.} 2013, \aap,
  553, A6

\bibitem[{{Jenkins}(2002)}]{2002ApJ...575..493J}
{Jenkins}, J.~M. 2002, \apj, 575, 493

\bibitem[{{Jofr{\'e}} {et~al.}(2018){Jofr{\'e}}, {Heiter}, \&
  {Soubiran}}]{2018arXiv181108041J}
{Jofr{\'e}}, P., {Heiter}, U., \& {Soubiran}, C. 2018, arXiv e-prints,
  arXiv:1811.08041

\bibitem[{{Kaeufl} {et~al.}(2004){Kaeufl}, {Ballester}, {Biereichel},
  {Delabre}, {Donaldson}, {Dorn}, {Fedrigo}, {Finger}, {Fischer}, {Franza},
  {Gojak}, {Huster}, {Jung}, {Lizon}, {Mehrgan}, {Meyer}, {Moorwood}, {Pirard},
  {Paufique}, {Pozna}, {Siebenmorgen}, {Silber}, {Stegmeier}, \&
  {Wegerer}}]{crires}
{Kaeufl}, H.-U., {Ballester}, P., {Biereichel}, P., {et~al.} 2004, in
  \procspie, ed. A.~F.~M. {Moorwood} \& M.~{Iye}, Vol. 5492, 1218--1227

\bibitem[{{Kurucz}(2005)}]{2005MSAIS...8...14K}
{Kurucz}, R.~L. 2005, Memorie della Societa Astronomica Italiana Supplementi,
  8, 14

\bibitem[{{Lindegren} \& {Dravins}(2003)}]{2003A&A...401.1185L}
{Lindegren}, L. \& {Dravins}, D. 2003, \aap, 401, 1185

\bibitem[{{Louden} \& {Wheatley}(2015)}]{Louden2015}
{Louden}, T. \& {Wheatley}, P.~J. 2015, \apjl, 814, L24

\bibitem[{{Ludwig} {et~al.}(2009){Ludwig}, {Caffau}, {Steffen}, {Freytag},
  {Bonifacio}, \& {Ku{\v c}inskas}}]{2009MmSAI..80..711L}
{Ludwig}, H., {Caffau}, E., {Steffen}, M., {et~al.} 2009, \memsai, 80, 711

\bibitem[{{Madhusudhan}(2012)}]{2012ApJ...758...36M}
{Madhusudhan}, N. 2012, \apj, 758, 36

\bibitem[{{Magic} \& {Asplund}(2014)}]{2014arXiv1405.7628M}
{Magic}, Z. \& {Asplund}, M. 2014, arXiv e-prints, arXiv:1405.7628

\bibitem[{{Magic} {et~al.}(2015){Magic}, {Chiavassa}, {Collet}, \&
  {Asplund}}]{2015A&A...573A..90M}
{Magic}, Z., {Chiavassa}, A., {Collet}, R., \& {Asplund}, M. 2015, \aap, 573,
  A90

\bibitem[{{Magic} {et~al.}(2013){Magic}, {Collet}, {Asplund}, {Trampedach},
  {Hayek}, {Chiavassa}, {Stein}, \& {Nordlund}}]{2013A&A...557A..26M}
{Magic}, Z., {Collet}, R., {Asplund}, M., {et~al.} 2013, \aap, 557, A26

\bibitem[{{Mihalas} {et~al.}(1988){Mihalas}, {Dappen}, \&
  {Hummer}}]{1988ApJ...331..815M}
{Mihalas}, D., {Dappen}, W., \& {Hummer}, D.~G. 1988, \apj, 331, 815

\bibitem[{{Miller-Ricci} {et~al.}(2009){Miller-Ricci}, {Seager}, \&
  {Sasselov}}]{2009ApJ...690.1056M}
{Miller-Ricci}, E., {Seager}, S., \& {Sasselov}, D. 2009, \apj, 690, 1056

\bibitem[{{Motalebi} {et~al.}(2015){Motalebi}, {Udry}, {Gillon}, {Lovis},
  {S{\'e}gransan}, {Buchhave}, {Demory}, {Malavolta}, {Dressing}, {Sasselov},
  {Rice}, {Charbonneau}, {Collier Cameron}, {Latham}, {Molinari}, {Pepe},
  {Affer}, {Bonomo}, {Cosentino}, {Dumusque}, {Figueira}, {Fiorenzano},
  {Gettel}, {Harutyunyan}, {Haywood}, {Johnson}, {Lopez}, {Lopez-Morales},
  {Mayor}, {Micela}, {Mortier}, {Nascimbeni}, {Philips}, {Piotto}, {Pollacco},
  {Queloz}, {Sozzetti}, {Vanderburg}, \& {Watson}}]{hd219134}
{Motalebi}, F., {Udry}, S., {Gillon}, M., {et~al.} 2015, \aap, 584, A72

\bibitem[{{Nordlund}(1982)}]{1982A&A...107....1N}
{Nordlund}, A. 1982, \aap, 107, 1

\bibitem[{{Nordlund} {et~al.}(2009){Nordlund}, {Stein}, \&
  {Asplund}}]{2009LRSP....6....2N}
{Nordlund}, {\AA}., {Stein}, R.~F., \& {Asplund}, M. 2009, Living Reviews in
  Solar Physics, 6, 2

\bibitem[{{Nortmann} {et~al.}(2018){Nortmann}, {Pall{\'e}}, {Salz},
  {Sanz-Forcada}, {Nagel}, {Alonso-Floriano}, {Czesla}, {Yan}, {Chen},
  {Snellen}, {Zechmeister}, {Schmitt}, {L{\'o}pez-Puertas}, {Casasayas-Barris},
  {Bauer}, {Amado}, {Caballero}, {Dreizler}, {Henning}, {Lamp{\'o}n}, {Montes},
  {Molaverdikhani}, {Quirrenbach}, {Reiners}, {Ribas}, {S{\'a}nchez-L{\'o}pez},
  {Schneider}, \& {Zapatero Osorio}}]{2018Sci...362.1388N}
{Nortmann}, L., {Pall{\'e}}, E., {Salz}, M., {et~al.} 2018, Science, 362, 1388

\bibitem[{{Oklop{\v{c}}i{\'c}} \& {Hirata}(2018)}]{oklopcic2018}
{Oklop{\v{c}}i{\'c}}, A. \& {Hirata}, C.~M. 2018, \apjl, 855, L11

\bibitem[{{Origlia} {et~al.}(2014){Origlia}, {Oliva}, {Baffa}, {Falcini},
  {Giani}, {Massi}, {Montegriffo}, {Sanna}, {Scuderi}, {Sozzi}, {Tozzi},
  {Carleo}, {Gratton}, {Ghinassi}, \& {Lodi}}]{giano2014}
{Origlia}, L., {Oliva}, E., {Baffa}, C., {et~al.} 2014, in \procspie, Vol.
  9147, 91471E

\bibitem[{{Piso} {et~al.}(2015){Piso}, {{\"O}berg}, {Birnstiel}, \&
  {Murray-Clay}}]{2015ApJ...815..109P}
{Piso}, A.-M.~A., {{\"O}berg}, K.~I., {Birnstiel}, T., \& {Murray-Clay}, R.~A.
  2015, \apj, 815, 109

\bibitem[{{Quirrenbach} {et~al.}(2014){Quirrenbach}, {Amado}, {Caballero},
  {Mundt}, {Reiners}, {Ribas}, {Seifert}, {Abril}, {Aceituno},
  {Alonso-Floriano}, {Ammler-von Eiff}, {Antona Jim{\'e}nez}, {Anwand
  -Heerwart}, {Azzaro}, {Bauer}, {Barrado}, {Becerril}, {B{\'e}jar},
  {Ben{\'\i}tez}, {Berdi{\~n}as}, {C{\'a}rdenas}, {Casal}, {Claret},
  {Colom{\'e}}, {Cort{\'e}s-Contreras}, {Czesla}, {Doellinger}, {Dreizler},
  {Feiz}, {Fern{\'a}ndez}, {Galad{\'\i}}, {G{\'a}lvez-Ortiz},
  {Garc{\'\i}a-Piquer}, {Garc{\'\i}a-Vargas}, {Garrido}, {Gesa}, {G{\'o}mez
  Galera}, {Gonz{\'a}lez {\'A}lvarez}, {Gonz{\'a}lez Hern{\'a}ndez},
  {Gr{\"o}zinger}, {Gu{\`a}rdia}, {Guenther}, {de Guindos},
  {Guti{\'e}rrez-Soto}, {Hagen}, {Hatzes}, {Hauschildt}, {Helmling}, {Henning},
  {Hermann}, {Hern{\'a}ndez Casta{\~n}o}, {Herrero}, {Hidalgo}, {Holgado},
  {Huber}, {Huber}, {Jeffers}, {Joergens}, {de Juan}, {Kehr}, {Klein},
  {K{\"u}rster}, {Lamert}, {Lalitha}, {Laun}, {Lemke}, {Lenzen}, {L{\'o}pez del
  Fresno}, {L{\'o}pez Mart{\'\i}}, {L{\'o}pez-Santiago}, {Mall}, {Mandel},
  {Mart{\'\i}n}, {Mart{\'\i}n-Ruiz}, {Mart{\'\i}nez-Rodr{\'\i}guez}, {Marvin},
  {Mathar}, {Mirabet}, {Montes}, {Morales Mu{\~n}oz}, {Moya}, {Naranjo},
  {Ofir}, {Oreiro}, {Pall{\'e}}, {Panduro}, {Passegger}, {P{\'e}rez-Calpena},
  {P{\'e}rez Medialdea}, {Perger}, {Pluto}, {Ram{\'o}n}, {Rebolo}, {Redondo},
  {Reffert}, {Reinhardt}, {Rhode}, {Rix}, {Rodler}, {Rodr{\'\i}guez},
  {Rodr{\'\i}guez-L{\'o}pez}, {Rodr{\'\i}guez-P{\'e}rez}, {Rohloff}, {Rosich},
  {S{\'a}nchez-Blanco}, {S{\'a}nchez Carrasco}, {Sanz-Forcada}, {Sarmiento},
  {Sch{\"a}fer}, {Schiller}, {Schmidt}, {Schmitt}, {Solano}, {Stahl}, {Storz},
  {St{\"u}rmer}, {Su{\'a}rez}, {Ulbrich}, {Veredas}, {Wagner}, {Winkler},
  {Zapatero Osorio}, {Zechmeister}, {Abell{\'a}n de Paco},
  {Anglada-Escud{\'e}}, {del Burgo}, {Klutsch}, {Lizon}, {L{\'o}pez-Morales},
  {Morales}, {Perryman}, {Tulloch}, \& {Xu}}]{carmenes2014}
{Quirrenbach}, A., {Amado}, P.~J., {Caballero}, J.~A., {et~al.} 2014, in
  \procspie, Vol. 9147, 91471F

\bibitem[{{Ram{\'{\i}}rez} {et~al.}(2008){Ram{\'{\i}}rez}, {Allende Prieto}, \&
  {Lambert}}]{2008A&A...492..841R}
{Ram{\'{\i}}rez}, I., {Allende Prieto}, C., \& {Lambert}, D.~L. 2008, \aap,
  492, 841

\bibitem[{{Schwarz} {et~al.}(2016){Schwarz}, {Ginski}, {de Kok}, {Snellen},
  {Brogi}, \& {Birkby}}]{Schwarz2016}
{Schwarz}, H., {Ginski}, C., {de Kok}, R.~J., {et~al.} 2016, \aap, 593, A74

\bibitem[{{Schwieterman} {et~al.}(2018){Schwieterman}, {Kiang}, {Parenteau},
  {Harman}, {DasSarma}, {Fisher}, {Arney}, {Hartnett}, {Reinhard}, {Olson},
  {Meadows}, {Cockell}, {Walker}, {Grenfell}, {Hegde}, {Rugheimer}, {Hu}, \&
  {Lyons}}]{2018AsBio..18..663S}
{Schwieterman}, E.~W., {Kiang}, N.~Y., {Parenteau}, M.~N., {et~al.} 2018,
  Astrobiology, 18, 663

\bibitem[{{Skartlien}(2000)}]{2000ApJ...536..465S}
{Skartlien}, R. 2000, \apj, 536, 465

\bibitem[{{Snellen} {et~al.}(2014){Snellen}, {Brandl}, {de Kok}, {Brogi},
  {Birkby}, \& {Schwarz}}]{2014Natur.509...63S}
{Snellen}, I.~A.~G., {Brandl}, B.~R., {de Kok}, R.~J., {et~al.} 2014, \nat,
  509, 63

\bibitem[{{Stempels} {et~al.}(2001){Stempels}, {Piskunov}, \&
  {Barklem}}]{2001ASPC..223..878S}
{Stempels}, H.~C., {Piskunov}, N., \& {Barklem}, P.~S. 2001, in Astronomical
  Society of the Pacific Conference Series, Vol. 223, 11th Cambridge Workshop
  on Cool Stars, Stellar Systems and the Sun, ed. R.~J. {Garcia Lopez},
  R.~{Rebolo}, \& M.~R. {Zapaterio Osorio}, 878

\bibitem[{{Trampedach} {et~al.}(2013){Trampedach}, {Asplund}, {Collet},
  {Nordlund}, \& {Stein}}]{2013ApJ...769...18T}
{Trampedach}, R., {Asplund}, M., {Collet}, R., {Nordlund}, {\AA}., \& {Stein},
  R.~F. 2013, \apj, 769, 18

\bibitem[{{Tremblay} {et~al.}(2013){Tremblay}, {Ludwig}, {Freytag}, {Steffen},
  \& {Caffau}}]{2013A&A...557A...7T}
{Tremblay}, P.~E., {Ludwig}, H.~G., {Freytag}, B., {Steffen}, M., \& {Caffau},
  E. 2013, \aap, 557, A7

\bibitem[{{Triaud} {et~al.}(2009){Triaud}, {Queloz}, {Bouchy}, {Moutou},
  {Collier Cameron}, {Claret}, {Barge}, {Benz}, {Deleuil}, {Guillot},
  {H{\'e}brard}, {Lecavelier Des {\'E}tangs}, {Lovis}, {Mayor}, {Pepe}, \&
  {Udry}}]{Triaud2009}
{Triaud}, A.~H.~M.~J., {Queloz}, D., {Bouchy}, F., {et~al.} 2009, \aap, 506,
  377

\bibitem[{{Wyttenbach} {et~al.}(2015){Wyttenbach}, {Ehrenreich}, {Lovis},
  {Udry}, \& {Pepe}}]{2015A&A...577A..62W}
{Wyttenbach}, A., {Ehrenreich}, D., {Lovis}, C., {Udry}, S., \& {Pepe}, F.
  2015, \aap, 577, A62

\bibitem[{{Zucker}(2003)}]{zucker03}
{Zucker}, S. 2003, \mnras, 342, 1291

\end{thebibliography}

\end{document}